\documentclass[twocolumn,showpacs,showkeys]{revtex4}

\usepackage{graphicx}
\usepackage{psfrag}
\usepackage{amssymb}
\usepackage{amsmath}
\usepackage{subfigure}
\usepackage{hyperref}

\newcommand{\dd}{\,\text{d}}                 % totale Differential:  \dd
\begin{document}
\title{Predictions from a stochastic polymer model for the MinDE dynamics in E.coli}
\author{Peter Borowski}
\email{borowski@math.ubc.ca}
\affiliation{Department of Mathematics, University of British Columbia, 1984 Mathematics Road, Vancouver, BC, V6T 1Z2 Canada; Pacific Institute for the Mathematical Sciences, Vancouver, Canada}
\author{Eric N. Cytrynbaum}
\email{cytryn@math.ubc.ca}
\affiliation{Department of Mathematics, University of British Columbia, 1984 Mathematics Road, Vancouver, BC, V6T 1Z2 Canada}

\date{\today}

\begin{abstract}
The spatiotemporal oscillations of the Min proteins in the bacterium Escherichia coli play an important role in cell division. A number of different models have been proposed to explain the dynamics from the underlying biochemistry. Here, we extend a previously described discrete polymer model from a deterministic to a stochastic formulation. We express the stochastic evolution of the oscillatory system as a map from the probability distribution of maximum polymer length in one period of the oscillation to the probability distribution of maximum polymer length half a period later and solve for the fixed point of the map with a combined analytical and numerical technique. This solution gives a theoretical prediction of the distributions of both lengths of the polar MinD zones and periods of oscillations -- both of which are experimentally measurable. The model provides an interesting example of a stochastic hybrid system that is, in some limits, analytically tractable.
\end{abstract}

\pacs{
%    {02.50.Ey}{Stochastic processes}   \and
%    {02.50.Fz}{Stochastic analysis}   \and
%    {82.40.Bj}{Oscillations, chaos, and bifurcations}   \and
%    {87.10.Ed}{Ordinary differential equations (ODE), partial differential equations (PDE), integrodifferential models}   \and
    {87.16.A-}{ Theory, modeling, and simulations}   
%    {87.16.Ka}{Filaments, microtubules, their networks, and supramolecular assemblies}   \and
    {87.17.Ee}{ Growth and division}  
    {87.10.Mn}{ Stochastic modeling} 
%{PACS-key}{discribing text of that key}   \and
%      {PACS-key}{discribing text of that key}
     } % end of PACS codes


\maketitle

\section{Introduction}

The spatiotemporal oscillations of the Min proteins in the bacterium Escherichia coli have been well studied both experimentally and theoretically (see~\cite{Lutkenhaus2009} for a recent review). Whereas many of the proposed models~\cite{Meinhardt2001,Meacci2005,Huang2003,Howard2001,Kerr2006,Fange2006,Kruse2002} describe the oscillations as a self-organised emergent property of a reaction-diffusion system, only a few~\cite{Drew2005,Pavin2006,Cytrynbaum2007,Tostevin2006} address the polymer nature of the relevant proteins that is evident in both in vitro~\cite{Hu2002,Suefuji2002} and in vivo~\cite{Shih2003,Szeto2005} experiments. Cytrynbaum and Marshall~\cite{Cytrynbaum2007} described the oscillatory dynamics of the Min proteins solely in terms of simple polymer assembly and disassembly dynamics coupled with concentration-dependent conditions for switching between these two states. Much of the experimentally observed behaviour of wildtype cells can be explained  within this framework as can a number of mutant-study observations. Here we analyze a stochastic version of the model introduced in~\cite{Cytrynbaum2007} and build a set of quantitative predictions that can be used to evaluate the model against data and results of other models.

Our stochastic model consists of a set of four interacting linear polymers, a pair of MinD and MinE polymers at either pole of the cell, each of which can be in either a growing or a shrinking state. For any fixed combination of states for the four polymers, the dynamics are deterministic and described by a system of ordinary differential equations for the lengths of the polymers. The full state of the system is thus determined by four discrete state variables (growing/shrinking) and four continuous variables (polymer lengths). Stochastic transitions between the discrete states are dependent on the cytoplasmic concentrations of the Min proteins and so indirectly on the polymer lengths.

Cytrynbaum and Marshall~\cite{Cytrynbaum2007} analysed a deterministic limit of infinitely high cooperativity, the solutions of which can be expressed easily in the form of fixed points of a one-dimensional map which we describe in the first subsection of Sec.~\ref{sec:results}. Formulating the model in this deterministic limit as a map is useful in understanding our approach to the full stochastic model which we focus on throughout the rest of Sec.~\ref{sec:results}. As with the deterministic model, the stochastic model can be reduced to a map, in this case one that takes the probability distribution for the maximum polymer length during a given period of the oscillation to the probability distribution for the same quantity half a period later. Our results include analysis of this probability distribution map and the calculation of its fixed point. In addition, from the analytical theory and numerical simulations, we calculate the distributions of other properties that are easily measured experimentally. This enables us to narrow down the parameter regime in which the results of our model agree with experimental observations. In particular, a high cooperativity ($n\approx 6$) in the nucleation of the MinE polymer is required. The cooperativity in nucleation of the MinD polymer turns out to be less crucial ($n\approx 3$ is sufficient). Also, over a wide range of parameter values, two stable solutions exist, one in which the Min proteins are entirely in the cytosol and one in which polymers form at either pole in an oscillatory manner. Stochastic transitions between these bistable states are more or less likely depending on parameter values. In the discussion, we put this finding into the context of recent experimental observations~\cite{Benjamin}, and compare our results to other modelling studies in the literature.

\section{Biology}

It is well established experimentally that two main processes control the position at which the rod-shaped bacterium E.coli divides. Nucleoid occlusion~\cite{Yu1999} prevents division at sites within close proximity of the nucleoid, which -- at the relevant time after DNA replication -- is everywhere in the cell except at the middle and near the two poles. The latter two sites are ruled out by the cooperating action of a group of three proteins, the Min proteins MinC, MinD, and MinE. In E.coli, a spatiotemporal oscillatory pattern formed by these proteins restricts division to the middle of the cell.

MinD is an ATPase that, in its ATP-bound form, binds to the inner cell membrane~\cite{deBoer1991}. MinE is found to activate the ATPase activity of membrane-bound MinD and thereby removes MinD from the membrane~\cite{Hu2001}. MinC co-localises to membrane-bound MinD~\cite{Raskin1999b} and is known to prevent assembly of FtsZ, one of the important players in forming the apparatus that constricts the cell during division~\cite{Hu1999}. Fluorescent labelling of MinD showed a preferred localisation to the polar regions of the membrane and away from midcell~\cite{Raskin1999}. This localisation appears to be the result of spatiotemporal oscillatory dynamics produced by the membrane-dependent interaction of MinD and MinE (for a review see~\cite{Lutkenhaus2009}). The still-uncertain mechanism by which this oscillation occurs is the motivation for our work.

The interaction of MinD and MinE in the presence of ATP and lipid membranes has been studied in vitro and MinD was found to accumulate into polymers and fibre bundles above certain concentrations~\cite{Suefuji2002}. In vivo, the fluorescently labelled Min proteins appear to be organised into helical structures on the inner wall of the cell membrane~\cite{Shih2003,Szeto2005}. The qualitative agreement between the in vitro and in vivo observation suggests that the Min proteins organise in the form of polymers in the cell, an assumption on which we base our theoretical model.

\section{Model -- a hybrid dynamical system}
\label{sec:model}

We assume that both MinD and MinE aggregate only in the form of polymers on the inner side of the cell's membrane. The Min oscillation results from an interplay between two MinD- and two MinE-polymers, one pair on each side of the cell. 

MinD monomers\footnote{To simplify the notation, we use the term monomer throughout this paper, even though the original model~\cite{Cytrynbaum2007} as well as experiments suggest a polymer formed of dimers.\label{footnote:monomers}} 
from the cytosol can start a polymer (i.e. nucleate) at one of the nucleation sites that are assumed to be positioned at each pole of the cell, an idea supported by recent experiments~\cite{Mileykovskaya2009,Mazor2008,Touhami2006}. The probability of nucleation on an empty site is proportional to the cytosolic MinD concentration raised to the power $n_{\mathrm{nuc}}$. We assume that each nucleation site, when occupied by a polymer, is incapable of nucleating a second polymer. The MinD polymer then elongates towards midcell at a rate proportional to the cytosolic MinD concentration. A MinE polymer can nucleate at the growing tip of the MinD polymer with a probability proportional to the cytosolic MinE concentration raised to the power $n_{\mathrm{cap}}$. It then grows backwards on top of the MinD polymer with a rate proportional to the MinE concentration. By inducing hydrolysis, the MinE subunits destabilise the underlying MinD subunits which disassemble from the tip, releasing both types of subunits into the cytosol. 

The geometry of our model cell is that of a cylinder with fixed length $L$ and diameter $2r$. Reported diffusion coefficients for the Min proteins are around $10\,\mathrm{\mu m^2s^{-1}}$~\cite{Meacci2006} so in a cell of length $2-3 \,\mathrm{\mu m}$ and with characteristic reaction times on the order of seconds, the cytosolic concentrations of MinD and MinE are essentially uniform throughout the cell. Also, the time scale of ADP-ATP exchange in cytosolic MinD is assumed to be fast compared to the oscillatory dynamics~\cite{Cytrynbaum2007}.

The equations governing the dynamics of the polymers are distinct for the different discrete states between which the system jumps. The system describing the behaviour therefore is a hybrid dynamical system -- a combination of continuous and discrete dynamics~\cite{Champneys:2008}. In our case, the continuous variables are four polymer-lengths $l_{l/r}^{\mathrm{D}}$ and $l_{l/r}^{\mathrm{E}}$ whose dynamics are determined by the values of the respective discrete state variables $S_{l/r}^{\mathrm{D}}$ and $S_{l/r}^{\mathrm{E}}$. In the following, we will provide the model equations and rephrase the model introduced in~\cite{Cytrynbaum2007} using a slightly different notation.

\subsection{Polymer dynamics}

For each of the four polymers (two MinD, two MinE), we track the projection of the polymers onto the long axis ($x$) of the cell (Fig.~\ref{fig:oscill_scheme}). The four variables $l^{\mathrm{D}}_l$, $l^{\mathrm{E}}_l$ , $l^{\mathrm{D}}_r$  and $l^{\mathrm{E}}_r$ describe these projected lengths for the polymers attached to the left and right pole of the cell, respectively. The relation between the full arc length of a helical polymer and its projection is given by $l/ \cos \theta$, where $\theta$ is the pitch of the helix. The parameter $\gamma = d\cos \theta$ is used to convert between the projected length $l$ of a polymer and the number of monomers it is made of (with monomer size $d$).

\subsubsection{MinD}
MinD polymerisation can start at either of two nucleation sites located at the two poles of the cell, i.e. at the positions $x=0$ and $x=L$. We assume that one end of the MinD-polymers is fixed to one of these stationary nucleation sites, whereas at the other end (the `tip'), the polymer can elongate or shorten. Growth and shrinkage are governed by the equations:

\begin{equation}
\frac{\dd}{\dd t}l_l^{\mathrm{D}} = \gamma k^{\mathrm{D}}(S_l^{\mathrm{D}}), \qquad \qquad \frac{\dd}{\dd t}l_r^{\mathrm{D}} = \gamma k^{\mathrm{D}}(S_r^{\mathrm{D}}) .
\label{eq.dyn_polymer_D}
\end{equation}
$k^{\mathrm{D}}(S_{l/r}^{\mathrm{D}})$ represents either constant disassembly or first order assembly and depends on the discrete state variable $S_{l/r}^{\mathrm{D}}$ of the polymer in question:

\begin{equation}
k^{\mathrm{D}}(S_{l/r}^{\mathrm{D}}) = \left\{ \begin{array}{cll} 
k_{\mathrm{on}}^{\mathrm{D}} c_{\mathrm{D}} & \text{ if } S_{l/r}^{\mathrm{D}}=1 & \text{(D-polymer growing)} \\
-k_{\mathrm{off}} & \text{ if } S_{l/r}^{\mathrm{D}}=0 & \text{(D-polymer shrinking)} \\
0 & \text{ if } S_{l/r}^{\mathrm{D}}=-1 & \text{(no D-polymer)}
 \end{array}\right.
\end{equation}

The dynamics of switching for the discrete state variables $S_{l/r}^{\mathrm{D}}$, between the three states $-1$, 0 and 1, will be explained in the next subsection.

The cytosolic concentration $c_{\mathrm{D}}$ of MinD is determined by conservation of monomers:
\begin{equation}
\frac{\lambda}{\gamma V}\left( l_l^{\mathrm{D}} + l_r^{\mathrm{D}}\right) + c_{\mathrm{D}} = c_{\mathrm{D,to}}
\label{eq.conserv_D}
\end{equation}
where $V=\pi r^2 L$ is the volume of the cell (in $\mathrm{\mu m^3}$), $\lambda\approx \frac{1}{602} \; \mathrm{\mu M \mu m^3}$ converts between particles per $\mathrm{\mu m^3}$ and $\mathrm{\mu M}$, and $ c_{\mathrm{D,to}}$ is the total concentration of MinD monomers.

The largest possible extension of a D-polymer is reached when all MinD is bound in one polymer. This is the case at
\begin{equation}
l_{\mathrm{max}}=\frac{\gamma V}{\lambda}c_{\mathrm{D,to}}.
\label{eq:lmax}
\end{equation}
For high total MinD concentrations, $l_{\mathrm{max}}$ can come close to $L$, i.e. the D-polymer would cover the whole cell from pole to pole. To avoid further assumptions on what happens if a polymer hits the opposite cell wall, we restrict ourselves here to total MinD concentrations that make these events very unlikely or impossible. With the parameters from Tab.~\ref{tab:parameters}, this means we consider maximal total MinD concentrations of around $5\;\mu\mathrm{M}$. In the simulations, the polymer simply stops growing in the unlikely case that it reaches the opposite cell wall.

\subsubsection{MinE}

We assume that the MinE polymer nucleates on the tip of the MinD polymer and grows on top of it towards the pole. The differential equation describing the projected length of the MinE polymer has the same simple structure as the one for the MinD polymer:
\begin{equation}
\frac{\dd}{\dd t}l_l^{\mathrm{E}} = \gamma k^{\mathrm{E}}(S_l^{\mathrm{E}}), \qquad \qquad  \frac{\dd}{\dd t}l_r^{\mathrm{E}} = \gamma k^{\mathrm{E}}(S_r^{\mathrm{E}}).
\label{eq.dyn_polymer_E}
\end{equation}
The same conversion factor $\gamma$ is used since we assume that MinE monomers bind to MinD monomers of the helix one-to-one. The MinE-polymer always starts growing from the non-polar tip of the MinD-polymer (Fig.~\ref{fig:oscill_scheme}). %The E-polymer then elongates on top of the D-polymer towards the cell pole.
The non-polar end of the MinDE-polymer falls off the membrane (with a slower speed than E-elongation) and disassembles. The state variable $S_{l/r}^{\mathrm{E}}$ defines the value of the growth rate as follows:

\begin{equation}
k^{\mathrm{E}}(S_{l/r}^{\mathrm{E}}) = \left\{ \begin{array}{cl} 
k_{\mathrm{on}}^{\mathrm{E}} c_{\mathrm{E}} - k_{\mathrm{off}} & \text{ if } S_{l/r}^{\mathrm{E}}=1 \text{ (E-polymer growing)} \\
-k_{\mathrm{off}} & \text{ if } S_{l/r}^{\mathrm{E}}=0 \text{ (E-polymer reached} \\
 & \text{ \qquad \qquad  cell wall } (l_{l/r}^{\mathrm{E}}=l_{l/r}^{\mathrm{D}})) \\
0 & \text{ if } S_{l/r}^{\mathrm{E}}=-1 \text{ (no E-polymer)}
 \end{array}\right.
\end{equation}

As for MinD, we assume the total number of MinE monomers to be constant:
\begin{equation}
\frac{\lambda}{\gamma V} \left( l_l^{\mathrm{E}} + l_r^{\mathrm{E}}\right)
 +  c_{\mathrm{E}}  = c_{\mathrm{E,to}} .
\label{eq.conserv_E}
\end{equation}

\subsection{Switching}
\label{ssec:model_switching}

We consider a stochastic description of switching between the three possible discrete states of the state variables $S_{l/r}^{\mathrm{D/E}}$ where the probability of  switching depends on the cytosolic concentrations of MinD/E. To capture the cooperative nature of the initiation of a polymer (e.g.~\cite{Mileykovskaya2003,Oosawa1975}), we assume that the instantaneous rates of nucleation and capping are proportional to a power of the respective cytosolic concentrations: $\lambda_{l/r}^{\mathrm{nuc}}(t)=k_{\mathrm{nuc}}c_{\mathrm{D}}^{n_{\mathrm{nuc}}}(t)$ and $\lambda_{l/r}^{\mathrm{cap}}(t)=k_{\mathrm{cap}}c_{\mathrm{E}}^{n_{\mathrm{cap}}}(t)$. $n_{\mathrm{nuc}}$ and $n_{\mathrm{cap}}$ are the cooperativities for the nucleation and the capping events, respectively. A D-polymer starts growing out of an empty nucleation site with probability $p_{\mathrm{nuc}}(t)=\lambda_{\mathrm{nuc}}(t)\dd t$ during the time interval $t..(t+\dd t)$ and a growing D-polymer gets capped (i.e. an E-polymer nucleates at its tip) with probability $p_{\mathrm{cap}}(t)=\lambda_{\mathrm{cap}}(t)\dd t$ during the time interval $t..(t+\dd t)$. Our main results involve calculating analytical expressions for the probability distributions when these switches happen. 

As a simplified model, we first consider the limit in which the cooperativities go to infinity. More specifically, we define the nucleation probability as $\lambda_{l/r}^{\mathrm{nuc}}(t) = \\ k_{\mathrm{nuc}}' ( c_{\mathrm{D}}(t)/c_{\mathrm{D},\mathrm{th}} )^{n_{\mathrm{nuc}}}$. In the limit $n_{\mathrm{nuc}}\rightarrow \infty$, the stochastic switching becomes deterministic with the nucleation event occurring as soon as $c_{\mathrm{D}}$ reaches the nucleation threshold $c_{\mathrm{D},\mathrm{th}}$. Capping is treated similarly.

 Note that this deterministic limit of our model is fundamentally different from the deterministic model of Drew et al. \cite{Drew2005} in that Drew et al. consider the mean field behavior of a population of filaments that switch between growing and shrinking states at average rates whereas we have individual filaments that switch at specific concentration-determined times. The difference between these two models is that the mean field model fails to admit oscillations without further biochemical assumptions (length-dependent growth speed), whereas the individual-filament model has an oscillatory solution over a large range of parameter values.

\subsection{Model summary}
\label{ssec:model_summary}

Eqs.~(\ref{eq.dyn_polymer_D})--(\ref{eq.conserv_E}) together with the above-mentioned switching dynamics of the discrete state variables $S_{l/r}^{\mathrm{D/E}}$ represent a hybrid dynamical system~\cite{Champneys:2008}. In the rest of this article, we analyse this system and present both analytical as well as numerical results that can be interpreted with regard to experimentally obtainable data.

A single MinD-polymer `life span' includes nucleation, growth, capping (nucleation of the MinE-polymer), and disassembly as described for the deterministic case in Tab.~\ref{tab:half_cycle}.

\begin{table*}
\begin{tabular}{l|l|c|c}
 & deterministic switching condition & $S^{\mathrm{D}}$ & $S^{\mathrm{E}}$ \\ \hline

(1) MinD polymer nucleates (nucleation event) & $c_{\mathrm{D}}(t)>c_{\mathrm{D,th}}$ & $-1 \rightarrow 1$ & \\

(2) MinE polymer nucleates (capping event) & $c_{\mathrm{E}}(t)>c_{\mathrm{E,th}}$ & $1\rightarrow 0$ & $-1 \rightarrow 1$ \\

(3) MinE polymer reaches cell wall & $l^{\mathrm{E}}=l^{\mathrm{D}}$ & & $1 \rightarrow 0$ \\

(4) MinDE polymer reaches cell wall & $l^{\mathrm{D}}=0$ & $0 \rightarrow -1$ & $0 \rightarrow -1$  
\end{tabular}
\caption{The series of states a MinD polymer on one pole goes through during one `life span'.}
\label{tab:half_cycle}
\end{table*}

\begin{figure*}
\centering
% helix_plot_04_paper.gnu plus xfig
\renewcommand{\thesubfigure}{}
\psfrag{ll}[c]{$l_l^{\mathrm{D}}=l_l^{\mathrm{E}}$}
\psfrag{lr}[c]{$l_r^{\mathrm{D}}$}
\psfrag{x=0}[c]{\scriptsize $x=0$}
\psfrag{x=L}[c]{\scriptsize $x=L$}
\psfrag{x}{$x$}
\subfigure[][\, \begin{tabular}{cccc } $S^{\mathrm{D}}_{l/r}$: &0 & / & 1 \\ $S^{\mathrm{E}}_{l/r}$: & 0 & / & -1\end{tabular}]{
\includegraphics[width=0.17\textwidth]{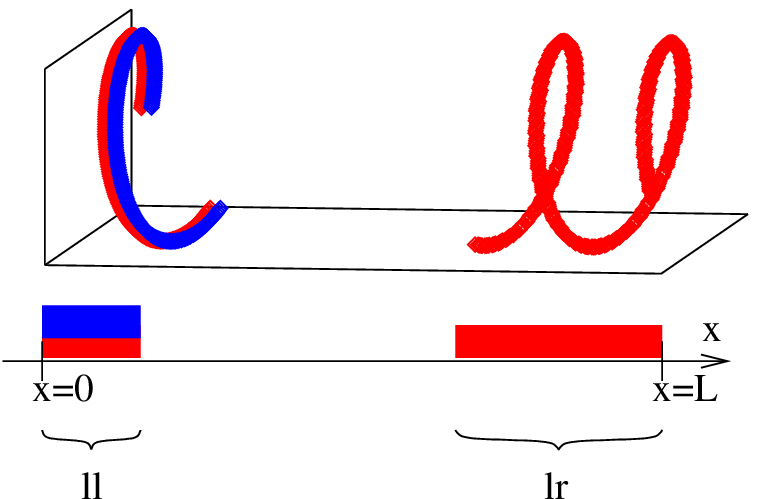}}
\subfigure[][\, \begin{tabular}{ccc} 0 & / & 0 \\ 0 & / & 1\end{tabular}]{
\includegraphics[width=0.17\textwidth]{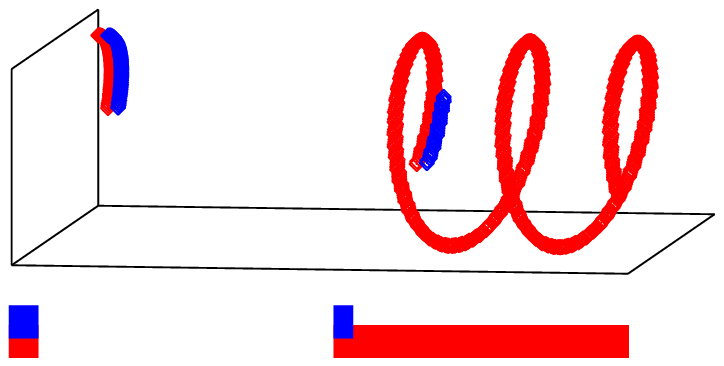}}
\subfigure[][\, \begin{tabular}{ccc}  -1 & / & 0 \\ -1 & / & 1\end{tabular}]{
\includegraphics[width=0.17\textwidth]{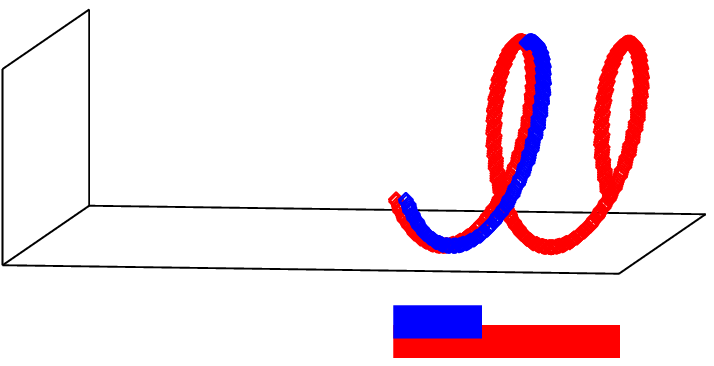}}
\subfigure[][\, \begin{tabular}{ccc} 1 & / & 0 \\  -1 & / & 1\end{tabular}]{
\includegraphics[width=0.17\textwidth]{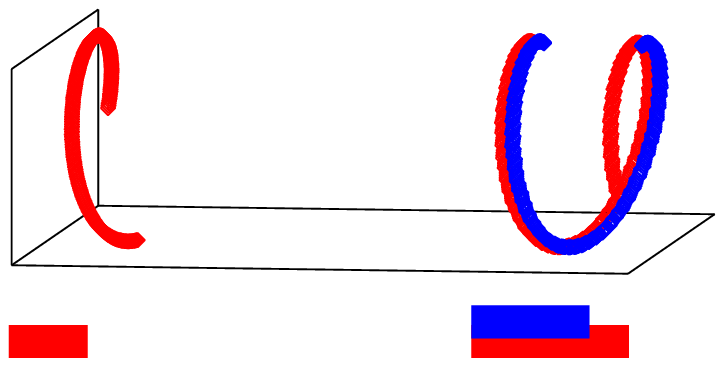}}
\subfigure[][\, \begin{tabular}{ccc}  1 & / & 0 \\  -1 & / & 0\end{tabular}]{
\includegraphics[width=0.17\textwidth]{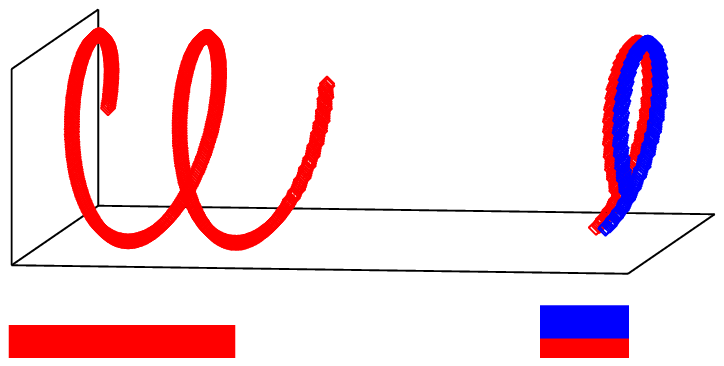}}
\caption{Scheme of the oscillations produced by the model. Shown are the two polymer helices (red -- MinD, blue -- MinE) on the left and the right side of the cell at different time points (time is increasing to the right). The values of the discrete state variables $S_l^{\mathrm{D}},S_l^{\mathrm{E}},S_r^{\mathrm{D}}$ and $S_r^{\mathrm{E}}$ are given under the figures.}
\label{fig:oscill_scheme}
\end{figure*}

\begin{figure}
\centering
\psfrag{Tc1}{$T^{c}_1$}
\psfrag{Td1}{$T^{d}_1$}
\psfrag{Tf2}{$T^{f}_2$}
\psfrag{Tn0}{$T^{n}_0$}
\psfrag{td0}{$t^d_0$}
\psfrag{tn1}{$t^n_1$}
\psfrag{tc1}{$t^c_1$}
\psfrag{tn2}{$t^n_2$}
\psfrag{ln0}{$l^{n}_0$}
\psfrag{lc1}{$l^{c}_1$}
\psfrag{lc2}{$l^{c}_2$}
\psfrag{ld1}{$l^{d}_1$}
\psfrag{ln1}{$l^{n}_1$}
\psfrag{td1}{$t^{d}_1$}
\psfrag{tn3}{$t^{n}_3$}
\psfrag{tc2}{$t^{c}_2$}
\psfrag{t}{$t$}
\psfrag{x}[c]{cell long axis $x\; [\mu\text{m}]$}
\psfrag{L=3}[c]{$L=3$}
\psfrag{1.5}[c]{1.5}
\psfrag{0}{0}
\psfrag{l}{$l$}
\psfrag{r}{$r$}
\psfrag{S=0}[c]{\, $S^{\mathrm{D}}=0$}
\psfrag{S=-1}[c]{\, $S^{\mathrm{D}}=-1$}
\psfrag{S=1}[c]{\, $S^{\mathrm{D}}=1$}
\includegraphics[width=.45\textwidth]{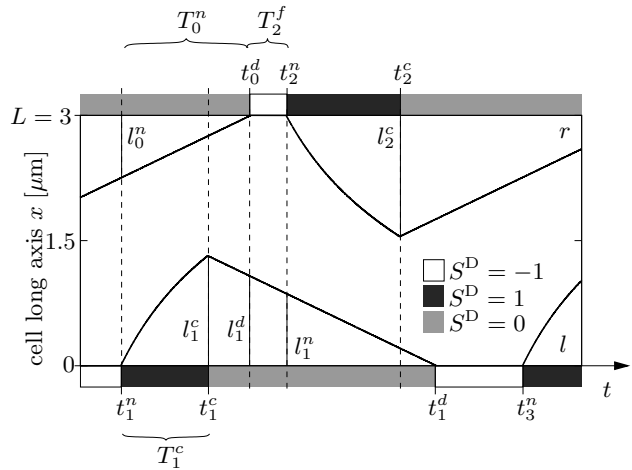}
\caption{A section of the regular oscillation pattern with stochastic switching to explain our notation. Shown is the temporal dynamics of the tip-positions ($x$) of the D-polymers as well as the value of the discrete state variables $S^{\mathrm{D}}_{l/r}$ (shaded areas). We use capital $T$ for time differences and little $t$ for time points. Even indices refer to the polymer anchored at the right pole of the cell and odd indices refer to the one anchored at the left pole.}
\label{fig:Ts_and_ls}
\end{figure}

The two poles can undergo alternating events of this type, which then constitutes an oscillatory solution. Such an oscillatory solution progresses in the following manner (cf. Fig.~\ref{fig:oscill_scheme} and Fig.~\ref{fig:Ts_and_ls} for notation). A MinD polymer capped (at time $t^c_1$) by a MinE polymer disassembles at one pole while the nucleation site at the other pole is either occupied by a disassembling MinDE polymer or remains empty. As the MinE polymer disassembles at the tip, it maintains a steady length by growing at its other end (treadmilling). Throughout this process, cytosolic MinD concentration increases thereby increasing the probability of MinD-polymer nucleation at the empty pole but cytosolic MinE concentration remains low. Once nucleation occurs (at time $t^n_2$), the nascent polymer grows towards midcell, depleting the cytosolic MinD pool. Because of the ordering of critical concentrations for nucleation and elongation, the existing MinE polymer elongates in preference to nucleation of a new MinE polymer capping the nascent MinD polymer. When the original MinD polymer reaches the same length as the steady-state-treadmilling MinE length, treadmilling is no longer possible and the cytosolic MinE concentration begins to rise. This raises the probability of capping the nascent MinD polymer (at time $t^c_2$). After capping, the system is back to the state we began describing but with the poles reversed. The first MinDE polymer completely disappears at time $t^d_1$.

For a specific subclass of the described oscillatory solution we are able to derive an analytical description for relevant probability distributions. This subclass we call {\it regular oscillations} and it is defined by $t^{c}_1 < t^{n}_{2} < t^{c}_2 < t^d_{1}$. A section of it is shown in Fig.~\ref{fig:Ts_and_ls}. This definition essentially means that the growing phase ($S^{\mathrm{D}}_{l/r}=1$) of one polymer falls completely within the shrinking phase of the other ($S^{\mathrm{D}}_{r/l}=0$).

Table~\ref{tab:parameters} gives an overview and numerical values for the parameters we used in this model. Most of them are taken in ranges reported in the experimental literature. Some are adjusted to values that lead to reasonable results of the model.

\begin{table}
\begin{tabular}{p{0.7cm}ccp{4cm}}
par. & value & unit & description  \\ \hline
% ************* adjust reference to footnote !!! ***********
$d$ & 2.5 & nm & increase in polymer length by addition of a single monomer
%diameter of a MinD monomer
 (see footnote on page~\pageref{footnote:monomers})~\cite{Suefuji2002} \\
$\theta$ & 1.4 & rad & pitch of the helical polymer~\cite{Shih2003} \\
$c_{\mathrm{D,to}}$ & 4 & $\mu$M & total MinD monomer conc. \\
$c_{\mathrm{E,to}}$ & 1.5 & $\mu$M & total MinE monomer conc. \\
$k_{\mathrm{off}}$ & 80 & $\mathrm{s^{-1}}$ & depolymerisation rate of the MinDE polymer \\
$k_{\mathrm{on}}^{\mathrm{D}}$ & 150 & $\mathrm{(\mu M \cdot s)^{-1}}$ & polymerisation rate of MinD on the membrane \\
$k_{\mathrm{on}}^{\mathrm{E}}$ & 320 & $\mathrm{(\mu M \cdot s)^{-1}}$ & polymerisation rate of MinE on the MinD polymer \\
$c_{\mathrm{D,th}}$ & 2.5 & $\mu$M & threshold conc. for the nucleation of a MinD polymer \\
$c_{\mathrm{E,th}}$ & 1.25 & $\mu$M & threshold conc. for the nucleation of a MinE polymer \\
$L$ & 3 & $\mu$m & length of cell \\
$r$ & 0.5 & $\mu$m & radius of cell \\
$k_{\mathrm{nuc}}$ & 0.015 & $\mathrm{s}^{-1}\mu \mathrm{M}^{-n_{\mathrm{nuc}}}$ & rate constant of nucleation \\
$k_{\mathrm{cap}}$ & 0.15 & $\mathrm{s}^{-1}\mu \mathrm{M}^{-n_{\mathrm{cap}}}$  & rate constant of capping \\
$n_{\mathrm{nuc}}$/ \,$n_{\mathrm{cap}}$ & 3--6 & & cooperativity of nucleation/capping \\
\end{tabular}
\caption{List of parameters used throughout this article. The concentrations $c_{\mathrm{D,to}}$ and $c_{\mathrm{E,to}}$ are consistent with values used in other modelling papers and are supported by experiment~\cite{Shih2002}. The rates governing polymer growth and decay are estimated from~\cite{Hale2001}. $L$ and $r$ are typical values seen in experiments and the other parameters are chosen such that the model produces reasonable results. A detailed discussion of the numerical values of some of the parameters can be found in~\cite{Cytrynbaum2007}.}
\label{tab:parameters}
\end{table}

%%%%%%%%%%%%%%%%%%%%%%%%%%%%%%%%%%%%%%%%%%%%%%%%%%%%%%%%%%%%%%%%%%%%%%%%%%%%%

\section{Results}
\label{sec:results}

A hybrid dynamical system must be solved piecewise and care must be taken at points of discontinuity which, in this case, occur each time a polymer switches state. Within the appropriate parameter ranges and for a regular oscillation (as defined above), we need only consider the progression through three of the six possible combinations of  discrete states of $(S^{\mathrm{D}}_{l},S^{\mathrm{D}}_{r})$ which occur in the sequence $(0,-1)\rightarrow(0,1)\rightarrow(0,0)$ during one half-period. The evolution of the continuous variable, for any given discrete state, requires a solution for disassembly and assembly of polymers. %EC
The solution of Eq.~\ref{eq.dyn_polymer_D} for a shrinking D-polymer ($S_{l/r}^{\mathrm{D}}=0$) is
\begin{equation}
l_{l/r}^{\mathrm{D}}(t) = l_{l/r}^{\mathrm{D}}(t_{l/r}^{c}) - \gamma k_{\mathrm{off}} (t-t_{l/r}^{c}),
\label{eq:decay}
\end{equation}
where $t_{l/r}^{c}$ is the time of the most recent capping of the left or right D-polymer, respectively.

The solution of Eq.~\ref{eq.dyn_polymer_D} for a growing D-polymer is dependent on the dynamic cytosolic concentration. During regular oscillations, a growing D-polymer only appears while the other D-polymer is disassembling (i.e., discrete state $(S^{\mathrm{D}}_{l/r},S^{\mathrm{D}}_{r/l})=(0,1)$). Assuming the polymer on the right is decaying, the cytosolic MinD concentration follows $c_{\mathrm{D}}(t)=c_{\mathrm{D,to}}-\frac{\lambda}{\gamma V}(l^{\mathrm{D}}_l(t)+l^{\mathrm{D}}_r(t^{c}_r)-\gamma k_{\mathrm{off}} (t-t^{c}_r))$. Substituting this into Eq.~\ref{eq.dyn_polymer_D} and solving the resulting equation with the initial condition $l^{\mathrm{D}}_l(t^{n}_l)=0$ ($t^{n}_l$ is the time at which the left polymer nucleates), one obtains
\begin{align}
l^{\mathrm{D}}_l(t) = & \frac{\gamma V}{\lambda}\left( c_{\mathrm{D,to}} - \frac{k_{\mathrm{off}}}{k_{\mathrm{on}}^{\mathrm{D}}}\right) - l^{\mathrm{D}}_r(t^{c}_r) + \gamma k_{\mathrm{off}} (t-t^{c}_r) \nonumber \\
 & - \left[ \frac{\gamma V}{\lambda}\left( c_{\mathrm{D,to}} - \frac{k_{\mathrm{off}}}{k_{\mathrm{on}}^{\mathrm{D}}}\right) - l^{\mathrm{D}}_r(t^{c}_r) + \gamma k_{\mathrm{off}} (t^{n}_l-t^{c}_r)\right] \nonumber \\
 & \; \cdot \exp\left[-\frac{\lambda k_{\mathrm{on}}^{\mathrm{D}}}{V}(t-t^{n}_l)\right].
\label{eq:lD_t}
\end{align}

Similar solutions can be found for the dynamics of the E-polymers. However, for the analytical treatment in this article, we restrict ourselves to the limit of fast E-ring formation: We assume that an E-polymer attains its steady state length right after its nucleation (i.e. after capping of the D-polymer).  Equating the rate of polymer decay at the medial end of the MinDE polymer with the growth rate of the MinE polymer gives the steady state E-polymer length $l_{\mathrm{E,ss}}=\frac{\gamma V}{\lambda}\left( c_{\mathrm{E,to}} - \frac{k_{\mathrm{off}}}{k_{\mathrm{on}}^{\mathrm{E}}}\right)$. With this simplification, the model reduces to a hybrid dynamical system with two continuous ($l^{\mathrm{D}}_{l/r}$) and two discrete ($S^{\mathrm{D}}_{l/r}$) variables. It is important to note that we make use of this approximation only for obtaining analytical results. In the numerical simulations we always model the full system with explicit E-polymer dynamics.% This leaves only the dynamics of switching to consider. 

For the rest of this article, we will drop the `D' superscript and denote the length of the D-polymer by $l$.

\subsection{Solution of the deterministic model}
\label{ssec:det}

As a basis for further discussions we briefly present the solution to the simplest version of the model described in the preceding section and previously addressed by Cytrynbaum and Marshall~\cite{Cytrynbaum2007}. We consider the case of deterministic switching (see Subsec.~\ref{ssec:model_switching}) and fast E-ring formation. The length of the MinD polymer at capping (the \emph{amplitude} of the oscillation) on one side can be expressed as a function of the capping length at the preceding capping on the other side: $l^{c}_{i+1} = f(l^{c}_{i})$ (cf. Fig.~\ref{fig:Ts_and_ls}). Thus, the problem of finding a periodic solution to the hybrid system is reduced to finding a fixed point of a one-dimensional map. Only in this simple version of the model such a one-dimensional map can be found, because only then the state of the system is completely determined by the length of only one of the polymers.

\begin{figure}
\centering
\psfrag{l1}{$l^{c}_{i}/L$}
\psfrag{l2}{$l^{c}_{i+1}/L$}
\subfigure[\,varying $c_{\mathrm{D,to}}$]{
\psfrag{one}[c]{\hspace{-.8cm} \scriptsize $c_{\mathrm{D,to}}=3\,\mu$M}
\psfrag{two}{\scriptsize 3.5}
\psfrag{thr}{\scriptsize 4}
\psfrag{fou}{\scriptsize 4.5}
\includegraphics[width=.48\textwidth]{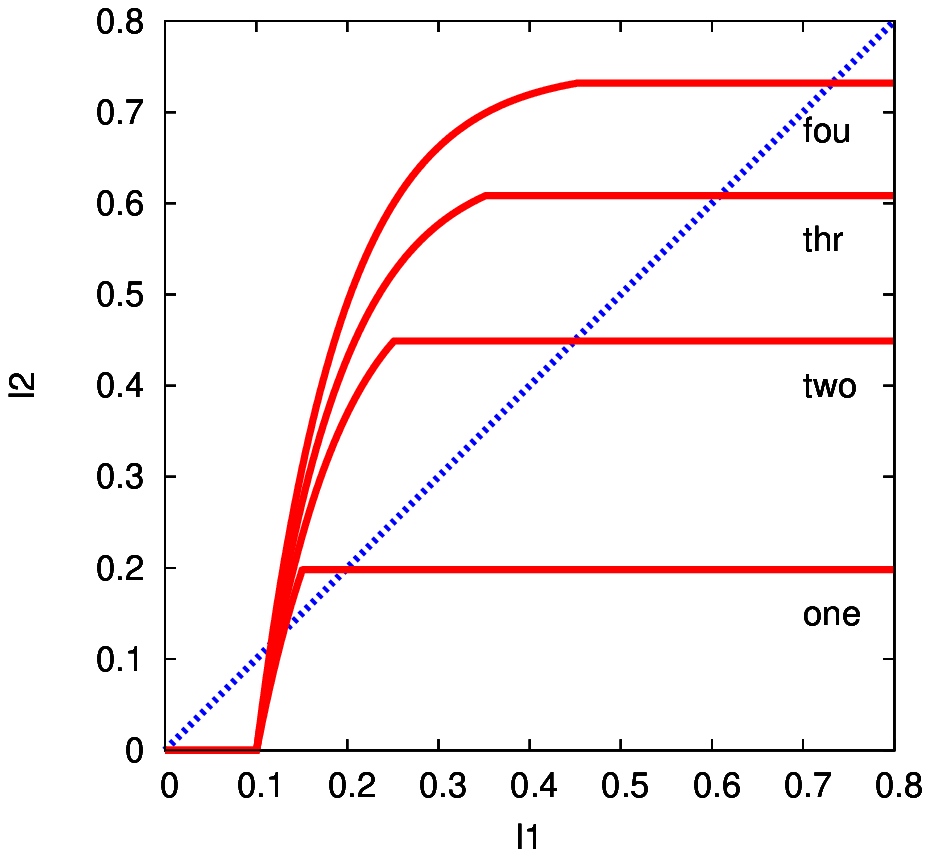}
\label{fig:mapD}
}
%\hfill 
\subfigure[\,varying $c_{\mathrm{E,to}}$]{
\psfrag{fo1}[c]{\hspace{-.8cm} \scriptsize $c_{\mathrm{E,to}}=1.3\,\mu$M}
\psfrag{th1}{\scriptsize 1.5}
\psfrag{tw1}{\scriptsize 1.7}
\psfrag{on1}{\scriptsize 1.9}
\includegraphics[width=.48\textwidth]{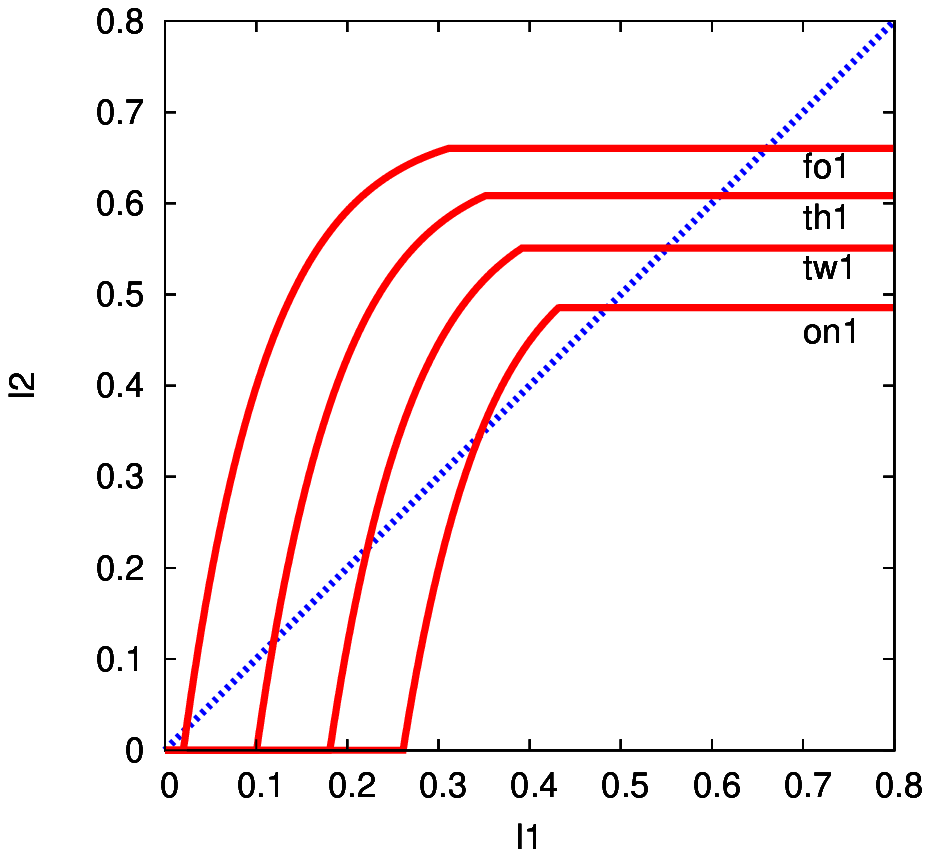}
\label{fig:mapE}
}
\caption{The dynamics of the deterministic version of the simplest model (infinitely fast E-ring formation) depicted in form of a discrete map. It displays the length of the MinD polymer relative to the cell length $L$ as a function of this lengths at the previous capping, i.e. the amplitude of the oscillation (see App.~\ref{app:map} for the equations). For intuitive analysis of the map, the identity line is added. Except for $c_{\mathrm{D,to}}$ and $c_{\mathrm{E,to}}$, standard parameters from Tab.~\ref{tab:parameters} were used.}
\label{fig:map}
\end{figure}

The calculation and the equations for the map are given in App.~\ref{app:map}. In Fig.~\ref{fig:map} the map is plotted for varying total concentrations of MinD and MinE.

The intersection of the map with $l^{c}_{i+1} = l^{c}_{i}$ shows three fixed
points which in terms of the hybrid system correspond to:
\begin{enumerate}
\item A stable cytosolic solution: There are no polymers ($l^{c}=0$) and all the Min proteins are in the cytosol as monomers.
\item An unstable oscillation (the map intersects the identity line with a slope larger than one).
\item Stable (slope is equal to zero) oscillations with constant amplitude (given in Eq.~\ref{eq:map_3}). 
\end{enumerate}
Depending on the initial conditions ($l_i(t=0)$), the system converges to one of the two stable states, i.e. each parameter set that allows for an oscillatory solution also includes a solution where no polymer exists (the cytosolic solution).

As can be seen from Fig.~\ref{fig:map}, the region of initial conditions that
lead to a stable oscillatory solution is biggest when $c_{\mathrm{D,to}}$ is
large and $c_{\mathrm{E,to}}$ is close to $c_{\mathrm{E,th}}$. If $c_{\mathrm{D,to}}$ is too small or $c_{\mathrm{E,to}}$ is too high, the map does not intersect with the identity line anymore and only the cytosolic solution remains. This corresponds qualitatively to the condition for the existence of oscillations as derived in~\cite{Cytrynbaum2007} for a simplified version of the model.

In the stochastic version of the model, low probability switching events can, under certain circumstances, lead to a transition between episodes of oscillations (with relatively constant amplitude) and almost purely cytosolic states. We will investigate this in the following sections and will refer to Fig.~\ref{fig:map} as the limiting case of infinite cooperativity.

%%%%%%%%%%%%%%%%%%%%%%%%%%%%%%%%%%%%%%%%%%%%%%%%%%%%%%%%%%%%%%%%%%%%%%%%%%%%%

\subsection{Probability distributions for experimentally measurable quantities}
\label{ssec:results_pdfs}

More realistic than the deterministic-threshold limit for nucleation and capping of the MinD polymer is stochastic switching that models the random nature of the underlying chemical reactions. From now on, we will use the stochastic rules for switching between the different states of polymers as introduced in Subsec.~\ref{ssec:model_switching}. In this section, we will characterise the changes such a stochastic switching rule introduces to the system. We will focus on two measures to characterise the robustness of the oscillatory solution of the stochastic model: the distribution of oscillation amplitudes and periods and the possibility of skipping beats. Both of these measures should be easy to obtain from experiments.

As long as the system undergoes regular oscillations, conditional probability distributions can be derived for the times at which nucleation and capping occur. Suppose the polymer on the right disappears at time $t^d$ (Fig.~\ref{fig:Ts_and_ls}) and, at that moment, the polymer on the left is shrinking and has length $l^d$. Given this, we denote the probability that a new polymer nucleates on the right at time $t$ by $P_{\mathrm{nuc}}(t|l^d)$. Thus, $P_{\mathrm{nuc}}(t|l^d) \dd t$ is the probability that nucleation on the right side happens in the time interval $t..(t+\dd t)$ where $t>t^d$. Similarly, if $t^n$ is the time of nucleation on the right and the polymer on the left is still shrinking and has length $l^n$, then $P_{\mathrm{cap}}(t|l^n) \dd t$ describes the probability for the capping of the growing polymer to happen in the time interval $t..(t+\dd t)$ after nucleation ($t>t^n$). Analytical expressions for these two probability distributions are computed in Apps.~\ref{app:nuc_distr} and~\ref{app:cap_distr} and plotted in Fig.~\ref{fig:Pcapt1}.

Using the two conditional probability distributions for nucleation and capping, analytical expressions for the steady state probability distribution of polymer lengths at capping and other relevant quantities can be derived. Similar to the description of the deterministic system in terms of a map (Subsec.~\ref{ssec:det}), we derive an integral relationship that maps the probability distribution for the polymer length $l^n$ (Fig.~\ref{fig:Ts_and_ls}) onto a new probability distribution of the same length half a period later. The derivation of this map $P(l^n_i)=F(P(l^n_{i-1}))$ and the expression for the map itself (Eq.~\ref{eq:Pln_iteration}) is presented in App.~\ref{app:map_prob}. The expression for the operator $F$ that relates two consecutive probability distributions involves complicated integrals and no general closed expression can be found. However, through numerical integration, an approximation to the steady state probability distribution can be computed iteratively. Typically, three or four iterates of the map $F$ (Eq.~\ref{eq:Pln_iteration}) are adequate for convergence with reasonable accuracy.

Only in the case of regular oscillations can the two stochastic processes, nucleation and capping, be separated and the probabilistic map (Eq.~\ref{eq:Pln_iteration}) be derived. Crucial for obtaining regular oscillations are high cooperativities in both capping and nucleation.

In the following, we present probability distributions for the amplitude and period of the oscillations obtained from numerical simulations\footnote{A simple Euler-forward routine (time step $10^{-4}$) was used to solve the differential equations. To reduce error in averages and to get smooth distributions, typically, simulations were run for $10^6 - 10^8$\,s of simulated time (i.e. roughly $2\cdot 10^4 - 2\cdot 10^6$ capping events).}. We compare these results to analytical results that we obtain by iterating the probabilistic map.

\subsubsection{Distribution of oscillation amplitude and dependence on cooperativities}

\begin{figure}
\centering
\subfigure[\, distribution of oscillation amplitudes]{
\psfrag{length}{$l^c\;[\mu\text{m}]$}
\psfrag{prob}{$p$}
\psfrag{key title}{$n_{\mathrm{nuc}}/n_{\mathrm{cap}}$}
\psfrag{n3c6}{3/6}
\psfrag{n3c3}{3/3}
\psfrag{n6c3}{6/3}
\psfrag{n6c6}{6/6}
\includegraphics[width=.48\textwidth]{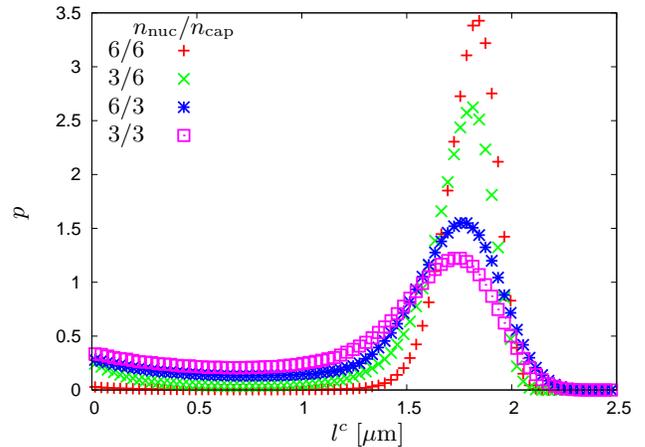}
\label{fig:distributions_numerically_amp}
}
\hfill 
\subfigure[\, distribution of periods]{
\psfrag{time}{$T\;[\text{s}]$}
\psfrag{prob}{$p$}
\psfrag{key title}{\hspace{-2mm}$n_{\mathrm{nuc}}/n_{\mathrm{cap}}$}
\psfrag{n3c6}{3/6}
\psfrag{n3c3}{3/3}
\psfrag{n6c3}{6/3}
\psfrag{n6c6}{6/6}
\includegraphics[width=.48\textwidth]{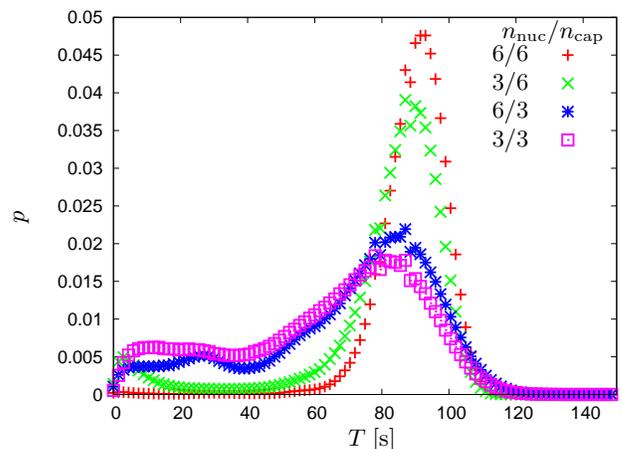}
\label{fig:distributions_numerically_T}
}
\caption{Numerically obtained distributions for oscillation amplitude (polymer lengths at capping) and periods (time between consecutive cappings on the same side) with varying cooperativity of nucleation (MinD) and capping. Standard parameters as of Tab.~\ref{tab:parameters} were used.% The lines are guides to the eye.
}
\label{fig:distributions_numerically}
\end{figure}

Fig.~\ref{fig:distributions_numerically} shows the results of long simulation runs of the full MinDE polymer model with stochastic switching. Plotted is the distribution of MinD-polymer lengths at capping (the amplitude of the oscillation) and the time between consecutive capping events on the same side (the period).

In experimental studies (both qualitative and quantitative~\cite{Shih2002}), large deviations in the oscillation amplitude and period are seldomly reported. For our model to reproduce this narrow distribution, high cooperativities in both nucleation and capping are crucial. Fig.~\ref{fig:distributions_numerically} shows that our model is especially sensitive to the cooperativity in capping: $n_{\mathrm{nuc}}=3$ still leads to a fairly narrow distribution, whereas both curves for $n_{\mathrm{cap}}=3$ have big contributions at unusually short capping lengths.

\begin{figure}
\centering
\subfigure[\, $n_{\mathrm{nuc}}=6,n_{\mathrm{cap}}=6$]{
\psfrag{length}{$l^c\;[\mu\text{m}]$}
\psfrag{prob}{$p$}
\label{fig:distr_sim_int_66}
\includegraphics[width=.231\textwidth]{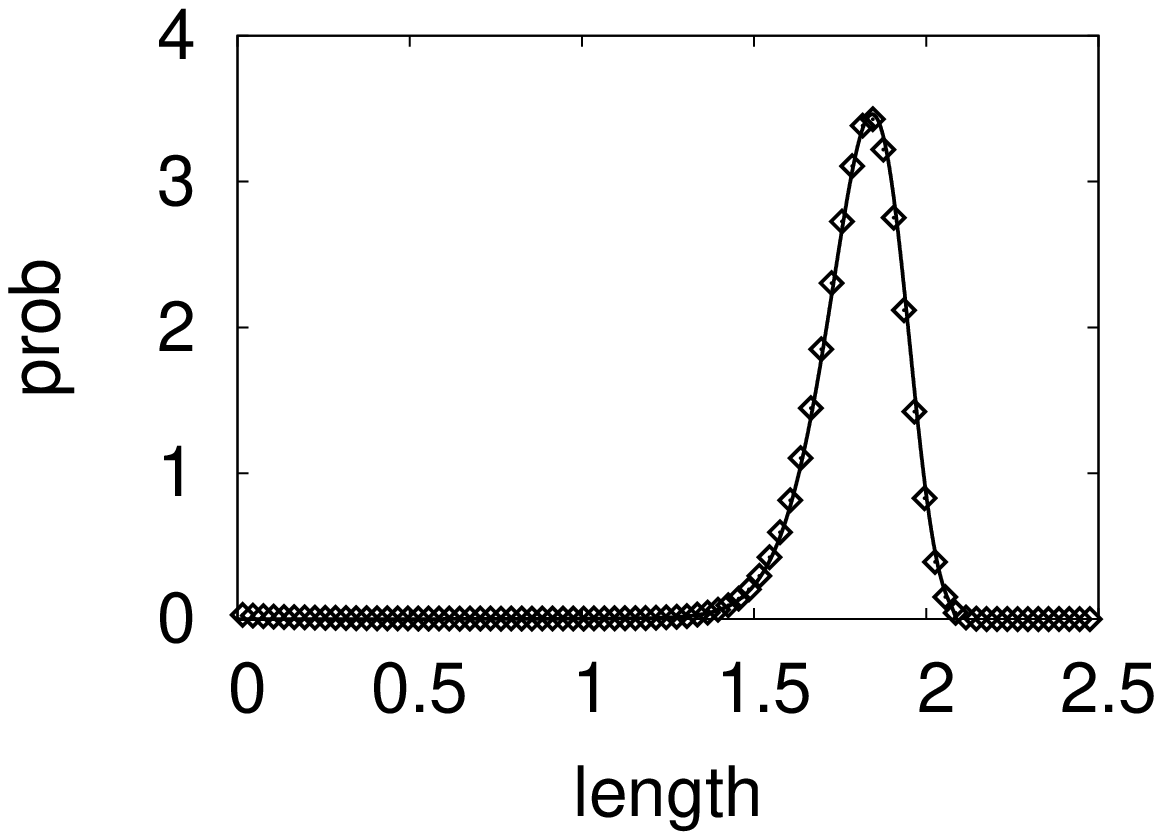}}
\hfill
\subfigure[\, $n_{\mathrm{nuc}}=3,n_{\mathrm{cap}}=6$]{
\psfrag{length}{$l^c\;[\mu\text{m}]$}
\psfrag{prob}{$p$}
\includegraphics[width=.231\textwidth]{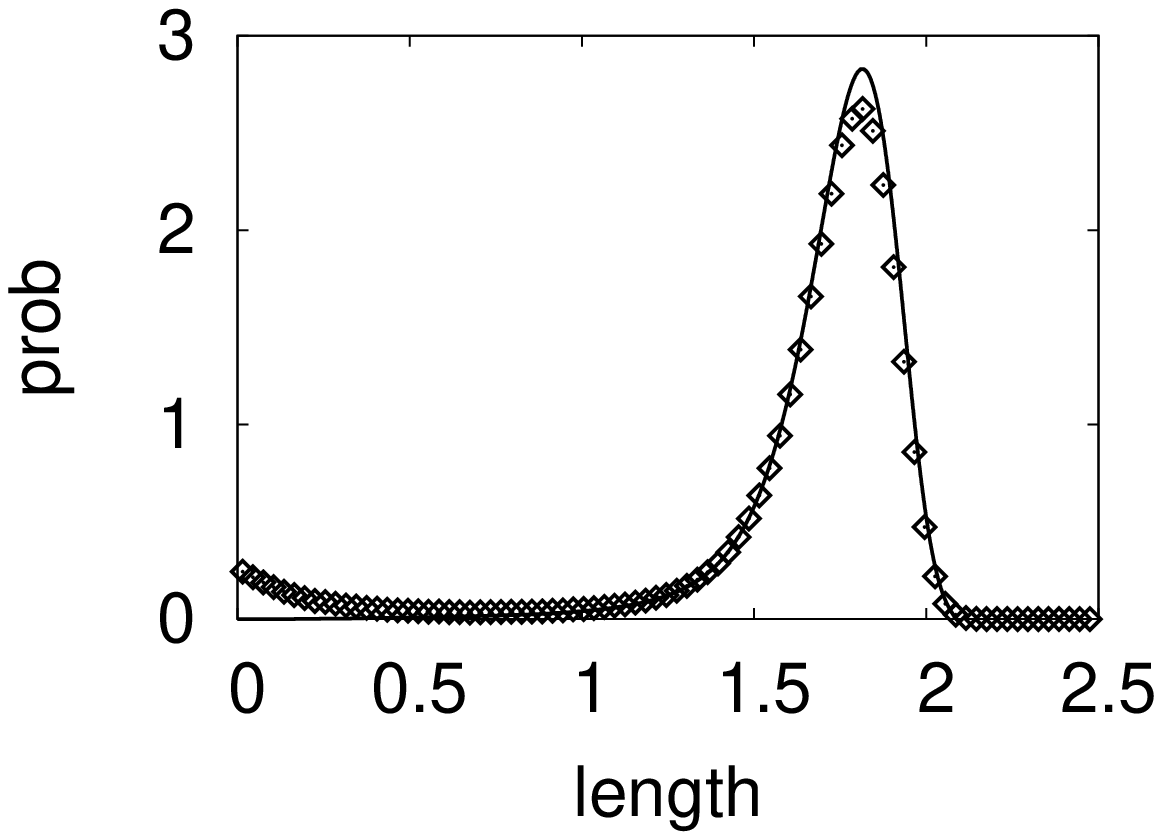}}
\hfill
\subfigure[\, $n_{\mathrm{nuc}}=6,n_{\mathrm{cap}}=3$]{
\psfrag{length}{$l^c\;[\mu\text{m}]$}
\psfrag{prob}{$p$}
\includegraphics[width=.231\textwidth]{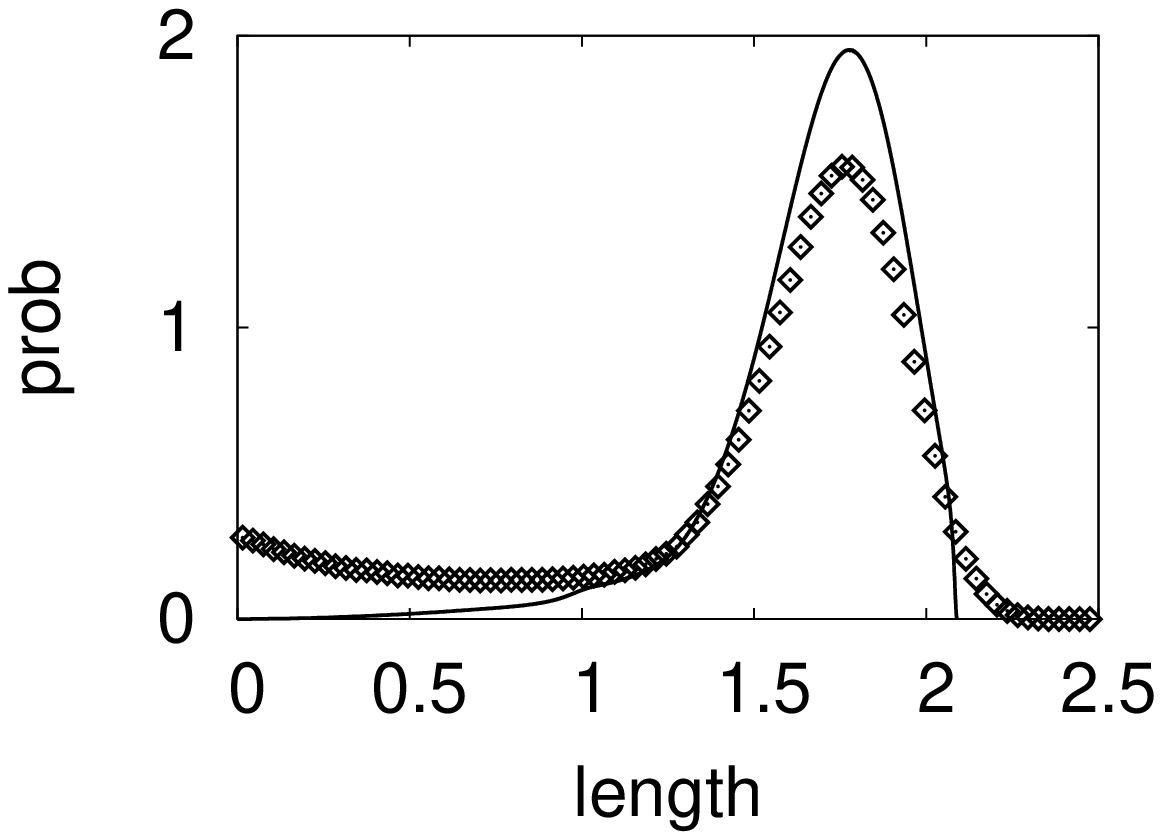}}
\hfill
\subfigure[\, $n_{\mathrm{nuc}}=3,n_{\mathrm{cap}}=3$]{
\psfrag{length}{$l^c\;[\mu\text{m}]$}
\psfrag{prob}{$p$}
\includegraphics[width=.231\textwidth]{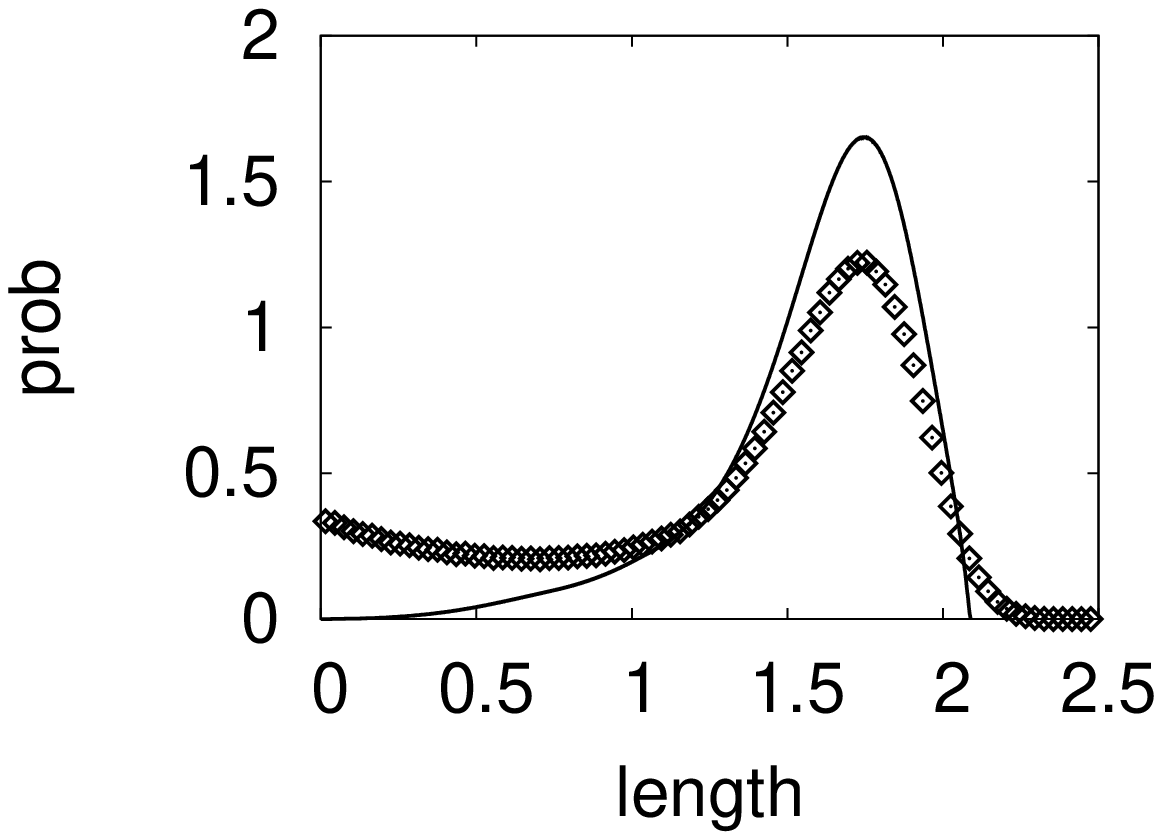}}
\caption{Comparison of the analytical result (line) with simulations (points -- same data as in Fig.~\ref{fig:distributions_numerically_amp}). For the analytical result, the probabilistic map (Eq.~\ref{eq:Pln_iteration}) was applied three consecutive times, starting with a uniform distribution. This steady state distribution in $l^n$ was then transformed into a distribution in $l^c$ by applying Eq.~\ref{eq:P(l_c)}.}
\label{fig:distr_sim_int}
\end{figure}

Fig.~\ref{fig:distr_sim_int} shows the same simulation results as of Fig.~\ref{fig:distributions_numerically_amp} together with the analytical result from the numerical iteration of the probabilistic map (Eqs.~\ref{eq:Pln_iteration} and~\ref{eq:P(l_c)}).

The plots confirm the applicability of our analytical solution as discussed above. For regular oscillations, the analytical result agrees well with the results from simulations. When cooperativities decrease, the increased probability at small polymer lengths is not accounted for by the analytical result. These are the events that are too far away from the stable oscillation point in Fig.~\ref{fig:map} and violate the regular oscillation assumption needed for the calculations in App.~\ref{app:map_prob}. Since both the numerical and the analytical distributions are normalised, the latter one shows higher values in the region corresponding to regular oscillations. The broader capping probability distribution in the case of lower $n_{\mathrm{cap}}$ (cf. also Fig.~\ref{fig:Pcapt1cap}) can lead to capping at short polymer lengths, which eventually can lead to a transition to a purely cytosolic state (see Subsec.~\ref{ssec:transitions}).

\subsubsection{Distribution of oscillation periods and dependence on total concentrations}

\begin{figure}
\centering
\subfigure[\, varying $c_{\mathrm{D,to}}$]{
\psfrag{key title}{\hspace{-2mm}$c_{\mathrm{D,to}}\;[\mu\text{M}]$}
\psfrag{time}{$T\;[\mathrm{s}]$}
\psfrag{prob}{$p$}
\includegraphics[width=.48\textwidth]{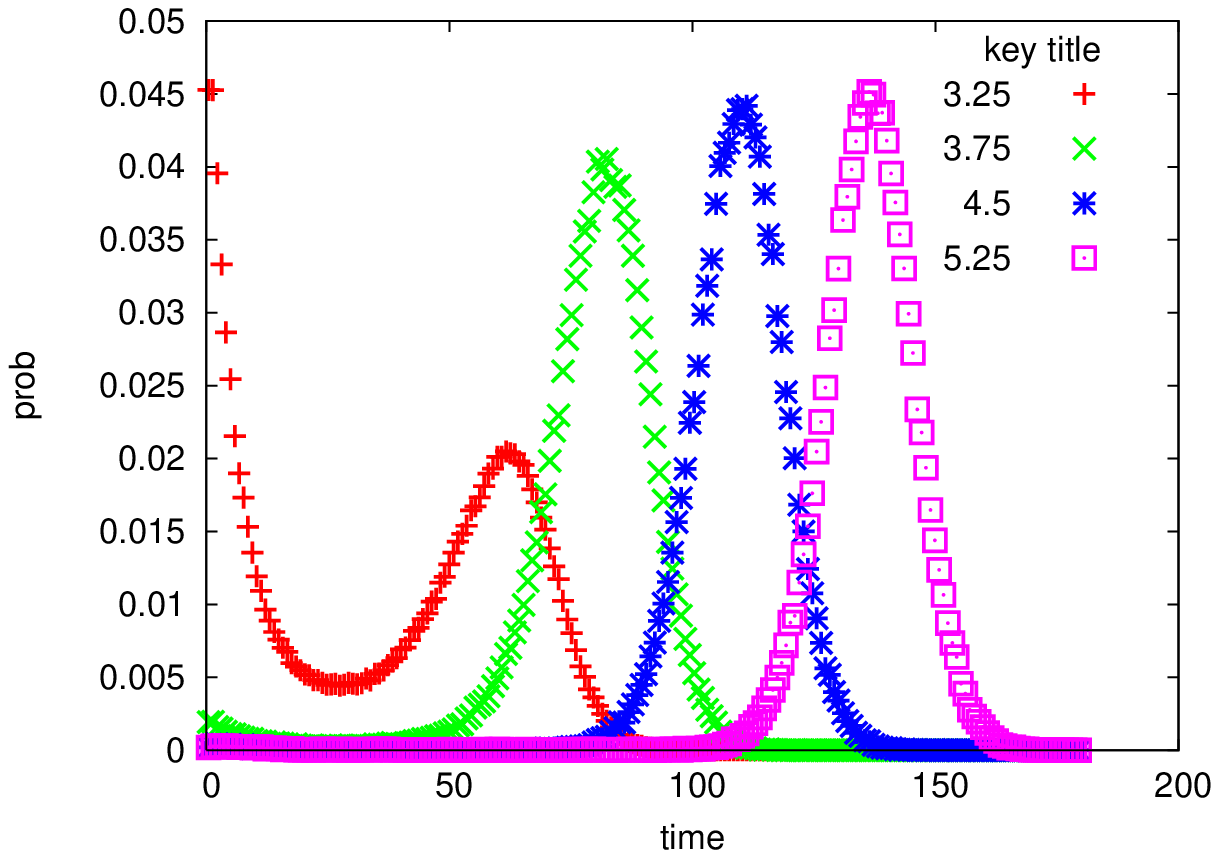}
\label{fig:T_hist_D}
}
\hfill
\subfigure[\, varying $c_{\mathrm{E,to}}$]{
\psfrag{key title}{\hspace{-2mm}$c_{\mathrm{E,to}}\;[\mu\text{M}]$}
\psfrag{time}{$T\;[\mathrm{s}]$}
\psfrag{prob}{$p$}
\includegraphics[width=.48\textwidth]{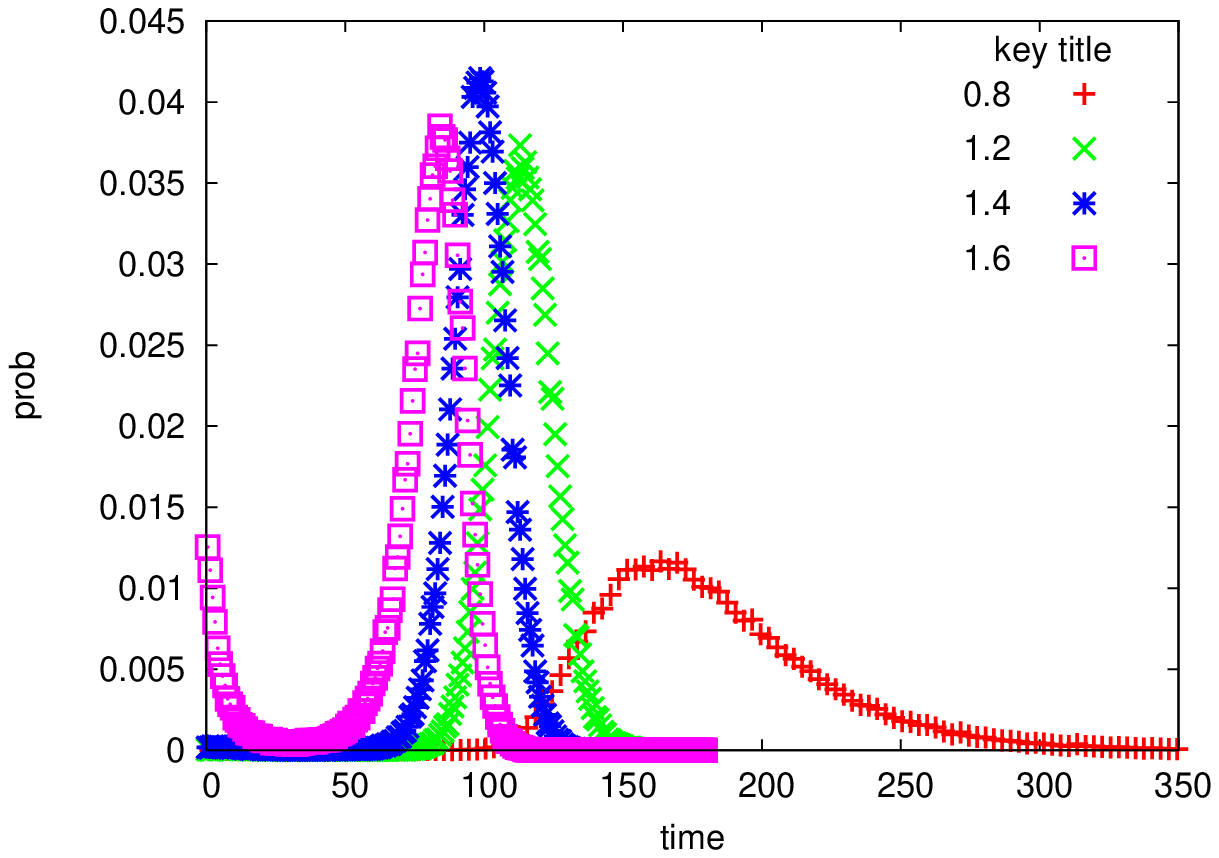}
\label{fig:T_hist_E}
}
\caption{Distributions of oscillation periods $T$ for different total concentrations $c_{\mathrm{D,to}}$ (upper panel) and $c_{\mathrm{E,to}}$ (lower panel). The data was obtained from simulations where the time between two consecutive cappings on the same side is considered to be $T$ (ignoring the dynamics on the other side). Both cooperativities (for nucleation and capping) are 6, $c_{\mathrm{E,to}}=1.5\,\mathrm{\mu M}$ (upper panel), and $c_{\mathrm{D,to}}=4\,\mathrm{\mu M}$ (lower panel).
}
\label{fig:T_hist}
\end{figure}

Another quantity that can be easily obtained from experiment is the period of the oscillations. In Fig.~\ref{fig:T_hist}, we plot the probability distributions of the period $T$ for different total concentrations of MinD and MinE as obtained from simulations.

At higher concentrations of MinD ($c_{\mathrm{D,to}}\gtrsim 3.75 \, \mu\text{M}$) and intermediate concentrations of MinE ($c_{\mathrm{E,to}} \approx 1.5 \,\mu\text{M}$), the distributions are sharply peaked around a single value of $T$ (see Fig.~\ref{fig:T_hist_D}). This parameter regime leads to regular oscillations. The amplitude of the oscillations decreases monotonically with $c_{\mathrm{D,to}}$, reflected in the leftward shift of the peak in Fig.~\ref{fig:T_hist_D}, consistent with the analogous feature in the deterministic version of the model (Fig.~\ref{fig:map}). The closer the amplitude comes to the unstable fixed point in Fig.~\ref{fig:map}, the more likely a D-polymer will be capped at a small length. This early capping at one pole leads to an earlier rise in MinE so that the subsequent capping event of the other pole occurs earlier as well. This positive feedback can drive the cell into the predominantly cytosolic state which appears in the stochastic version of the model as a second peak close to $T=0$ (Fig.~\ref{fig:T_hist_D}, $c_{\mathrm{D,to}}= 3.25 \, \mu\text{M}$).

A similar distribution is obtained if the total MinE concentration is chosen too high (shown in Fig.~\ref{fig:T_hist_E}, $c_{\mathrm{E,to}}= 1.6 \, \mu\text{M}$) with capping of the D-polymers occurring quite early. For low concentrations of MinE, the instantaneous capping rate is reduced and capping of the D-polymer can occur late (cf. also Fig.~\ref{fig:Pcapt1cap}). In extreme cases ($c_{\mathrm{E,to}}=0.8 \, \mu\text{M}$) the D-polymer can reach maximum length $l_{\mathrm{max}}$ (Eq.~\ref{eq:lmax}). Accordingly, the distribution of periods is shifted towards higher $T$ and tails out slowly.

\begin{figure}
\centering
\subfigure[\, $3.25\;\mu\mathrm{M}/1.5\;\mu\mathrm{M}$]{
\psfrag{T}{$T\;[\mathrm{s}]$}
\psfrag{prob}{$p$}
\label{fig:distr_sim_int_c_lowD}
\includegraphics[width=.231\textwidth]{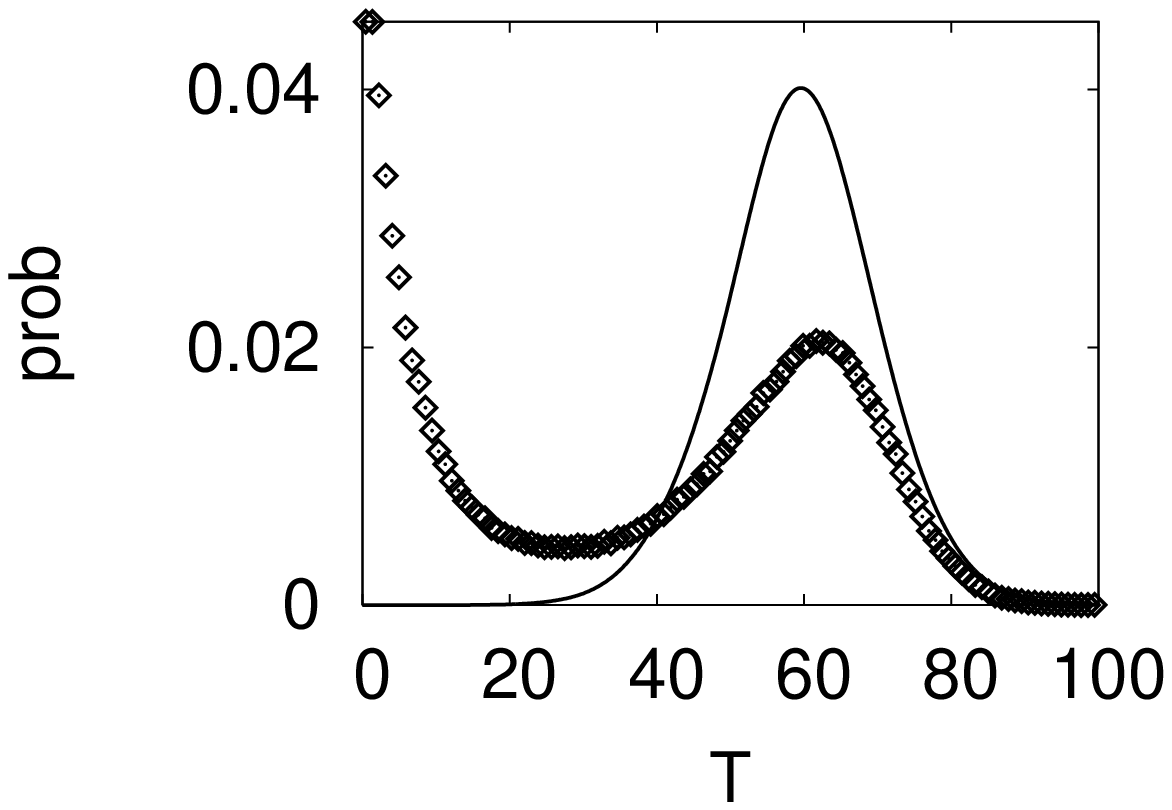}}
\hfill
\subfigure[\,  $4\;\mu\mathrm{M}/1.5\;\mu\mathrm{M}$]{
\psfrag{T}{$T\;[\mathrm{s}]$}
\psfrag{prob}{$p$}
\label{fig:distr_sim_int_c_standard}
\includegraphics[width=.231\textwidth]{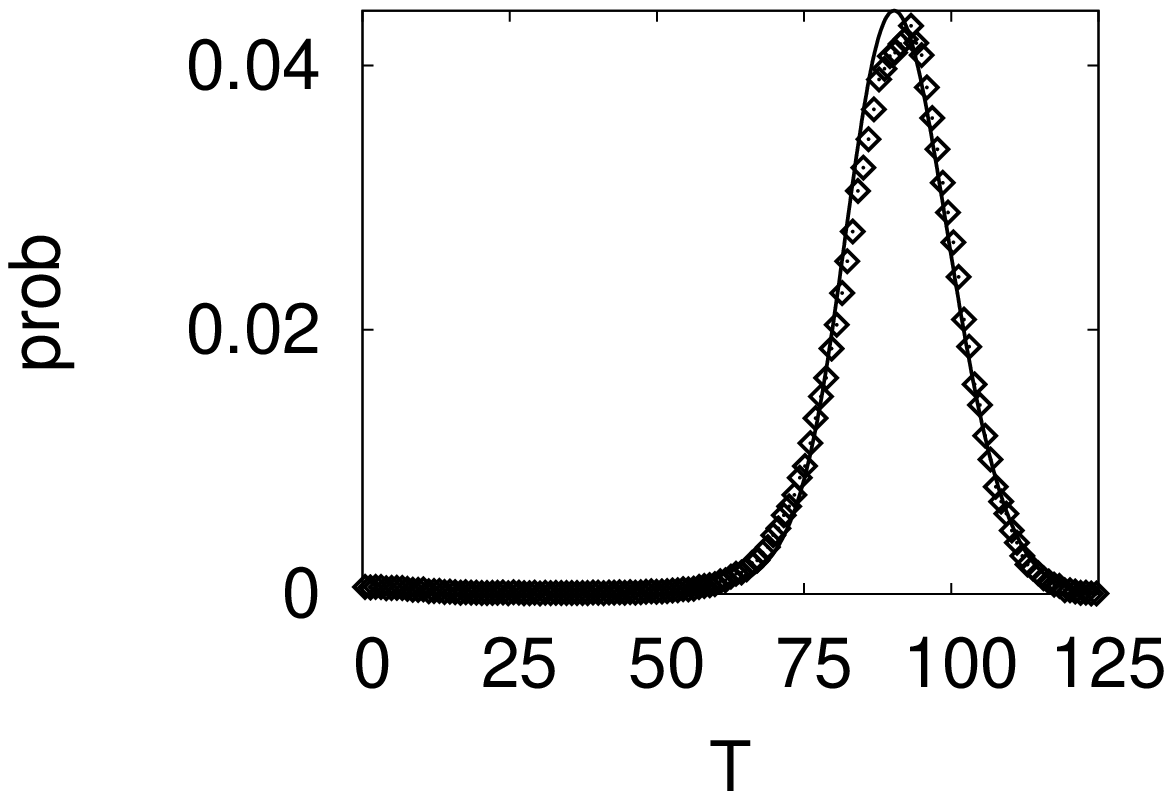}}
\hfill
\subfigure[\,  $4\;\mu\mathrm{M}/0.8\;\mu\mathrm{M}$]{
\psfrag{T}{$T\;[\mathrm{s}]$}
\psfrag{prob}{$p$}
\label{fig:distr_sim_int_c_lowE}
\includegraphics[width=.231\textwidth]{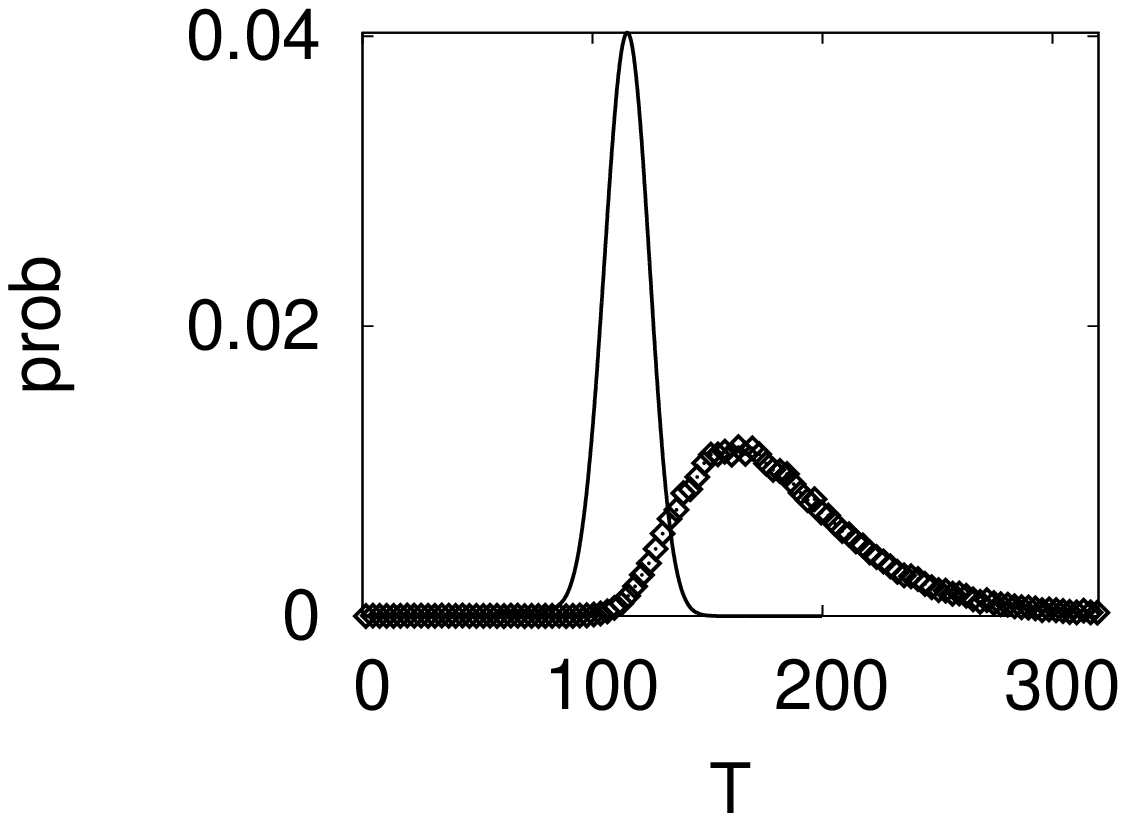}}
\hfill
\subfigure[\,  $4\;\mu\mathrm{M}/1.6\;\mu\mathrm{M}$]{
\psfrag{T}{$T\;[\mathrm{s}]$}
\psfrag{prob}{$p$}
\label{fig:distr_sim_int_c_highE}
\includegraphics[width=.231\textwidth]{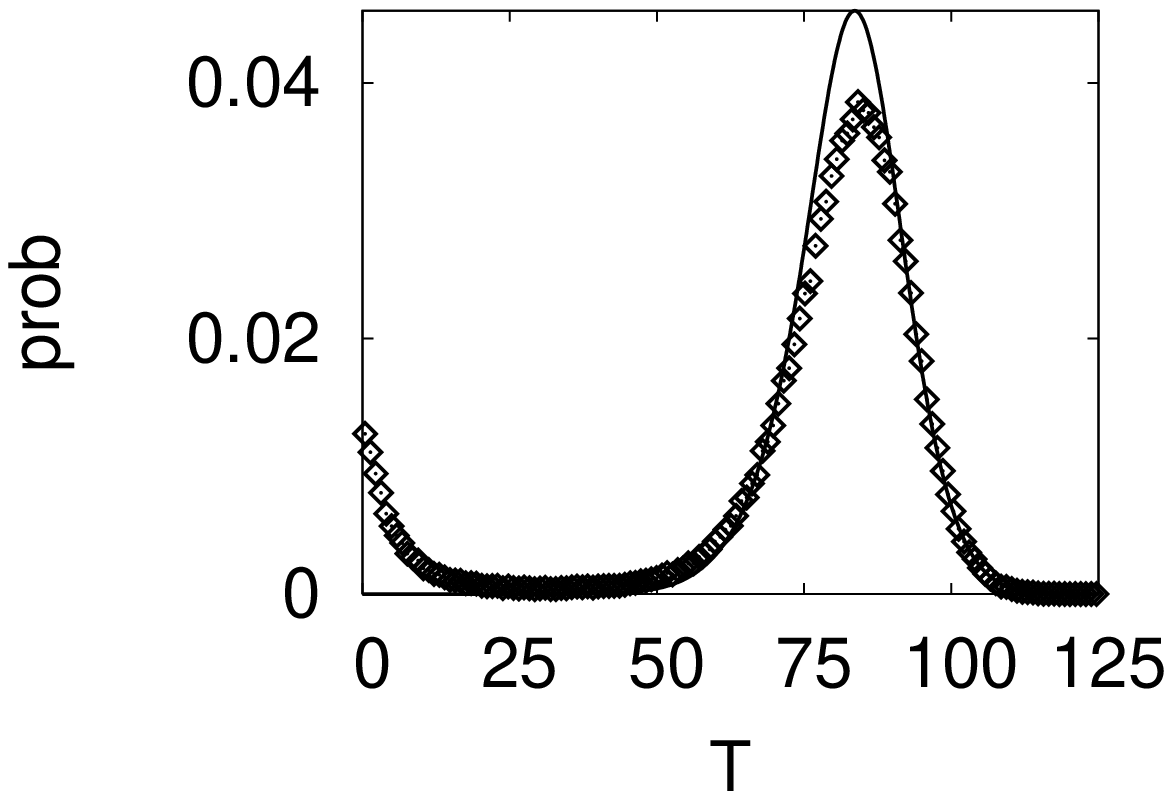}}
\caption{Comparison of the analytical result (line) with simulations (points). For the analytical result, the probabilistic map (Eq.~\ref{eq:Pln_iteration}) was applied three consecutive times, starting with a uniform distribution. This steady state distribution in $l^n$ was then transformed into a distribution in the period $T$ by applying Eqs.~\ref{eq:p(T_h)} and~\ref{eq:p(T)}. The simulation data in (a), (c), and (d) is from Fig.~\ref{fig:T_hist}, the data in (b) from Fig.~\ref{fig:distributions_numerically}. The numbers in the subcaptions are $c_{\mathrm{D,to}}/c_{\mathrm{E,to}}$.}
\label{fig:distr_sim_int_c}
\end{figure}

Using the steady state probability distribution for $l^n$, an analytical integral expression for the probability distribution of the period $T$ can be derived for the case of regular oscillations (see  App.~\ref{app:pdf_T}). Fig.~\ref{fig:distr_sim_int_c} shows a comparison of this analytical result with simulation data. Since the analytical description only contains the regular oscillations, it fails to produce the peaks at short periods in Fig.~\ref{fig:distr_sim_int_c_lowD} and~\ref{fig:distr_sim_int_c_highE}. Also, low total concentrations of MinE (e.g., $c_{\mathrm{E,to}}\lesssim 1.2\,\mu\mathrm{M}$) lead to an irregular oscillation regime by violating the scheme shown in Fig.~\ref{fig:Ts_and_ls} in that capping of a new polymer is likely to occur after complete disassembly of the existing polymer at the other pole (through $t^c_{i}>t^d_{i-1}$). Accordingly, the analytical solution fails to give a good approximation (Fig.~\ref{fig:distr_sim_int_c_lowE}). In Fig.~\ref{fig:distr_sim_int_c_standard}, the standard parameter set (Tab.~\ref{tab:parameters}) is used and analytical  and numerical results agree. Fig.~\ref{fig:distr_sim_int_c_standard} and~\ref{fig:distr_sim_int_66} represent the parameter values that provide the closest match to the experimental observations of sustained and regular oscillations.

\begin{figure}
\centering
%\subfigure[\, varying $c_{\mathrm{D,to}}$; $c_{\mathrm{E,to}}=1.5\;\mu\text{M}$ ]{
\psfrag{cDto}[c]{$c_{\mathrm{D,to}}\;[\mu\text{M}]$}
\psfrag{mean T}{$\bar{T}\;[\text{s}]$}
\includegraphics[width=.4\textwidth]{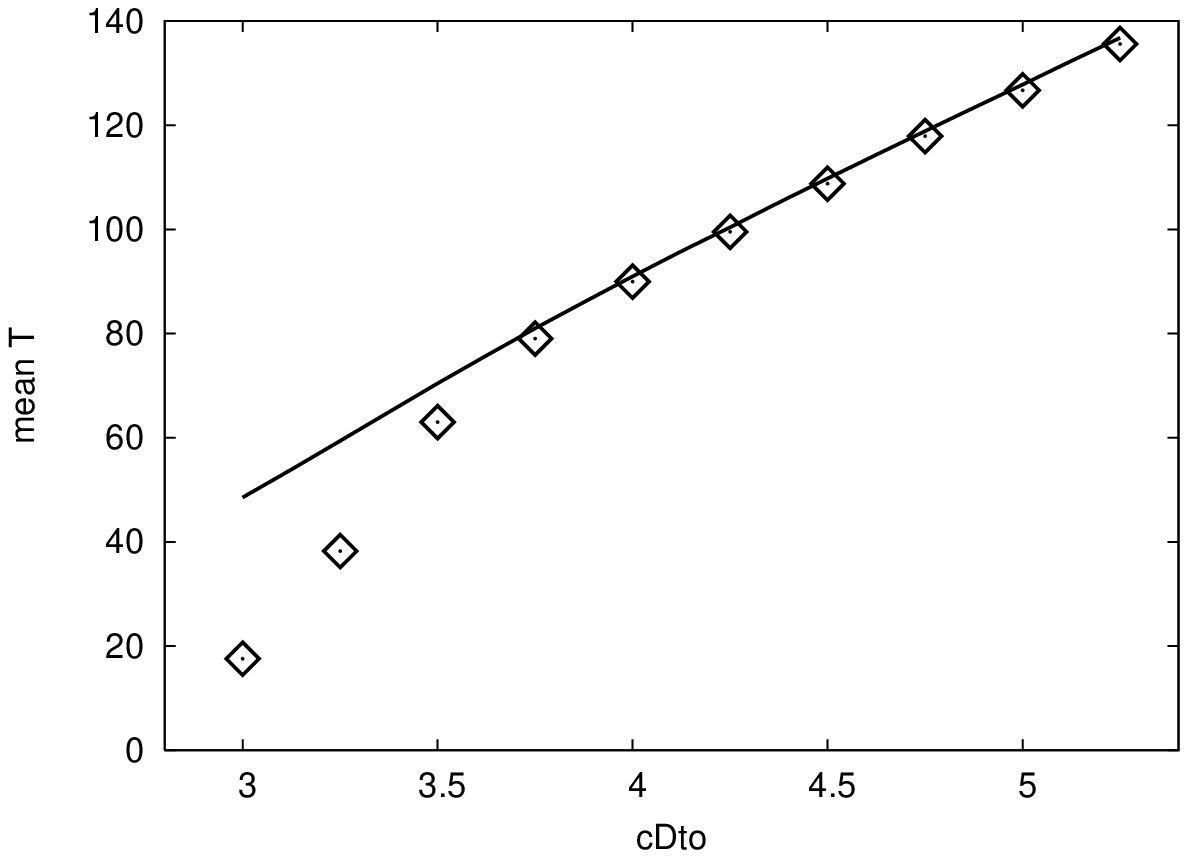}
%}
%\hfill
%\subfigure[\, varying $c_{\mathrm{E,to}}$; $c_{\mathrm{D,to}}=4\;\mu\text{M}$]{
\psfrag{cEto}[c]{$c_{\mathrm{E,to}}\;[\mu\text{M}]$}
\psfrag{mean T}{$\bar{T}\;[\text{s}]$}
\includegraphics[width=.4\textwidth]{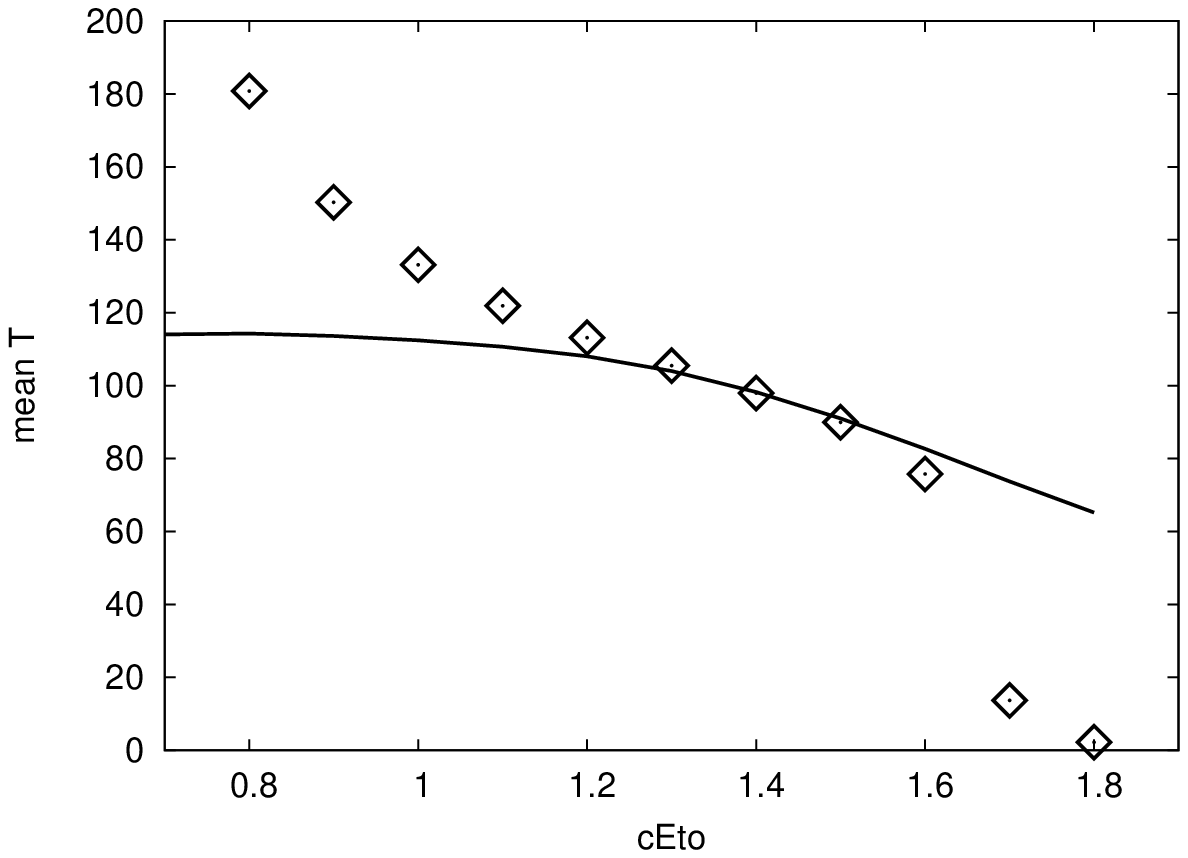}
%}
\caption{Mean oscillation period $\bar{T}$ for varying total concentrations of the two Min proteins. Both the analytical result (line) as well as results from simulations (points) are shown. Cooperativities for nucleation and capping are both 6, $c_{\mathrm{E,to}}=1.5\;\mu\text{M}$ in the upper panel and $c_{\mathrm{D,to}}=4\;\mu\text{M}$ in the lower.}
\label{fig:mean_T}
\end{figure}

To illustrate the range of validity of our analytical calculation, we compare the means of the numerically and analytically obtained probability distributions in Fig.~\ref{fig:mean_T}.

The deviations in mean period in Fig.~\ref{fig:mean_T} reiterate that the analytic expression is valid provided MinD concentration is high and MinE concentration is intermediate, ensuring regular oscillations. For small $c_{\mathrm{D,to}}$, the analytical result is always an overestimate of the numerical result, since the analytical treatment does not capture the second peak in the distribution at short $T$ (Fig.~\ref{fig:distr_sim_int_c_lowD}). The same overestimate results for high $c_{\mathrm{E,to}}$ (Fig.~\ref{fig:distr_sim_int_c_highE}), whereas for low $c_{\mathrm{E,to}}$, the analytical calculation provides an underestimate of the period. Here, the long tail of the numerically obtained distributions is not covered (Fig.~\ref{fig:distr_sim_int_c_lowE}).

Our results as shown in Fig.~\ref{fig:mean_T} agree qualitatively with experiments. Overexpression of MinD has been shown to increase the oscillation period~\cite{Raskin1999b} and overexpression of MinE is known to disrupt the normal division placement~\cite{deBoer1989}, probably due to destroying the oscillatory pattern (which corresponds to our numerical solution of $\bar{T}\approx 0\,\mathrm{s}$ for high MinE concentration). In the deterministic version of the model, this corresponds to the deterministic map not intersecting with the identity line anymore (Fig.~\ref{fig:mapE}).

\subsubsection{Beat-skipping}

\begin{figure}
\centering
%\subfigure[\, varying $c_{\mathrm{D,to}}$]{
\psfrag{cDto}{$c_{\mathrm{D,to}}\;[\mu\text{M}]$}
\psfrag{fraction}[c]{fraction of skipped beats}
\psfrag{key title}{$n_{\mathrm{nuc}}/n_{\mathrm{cap}}$}
\psfrag{n3c6}{3/6}
\psfrag{n3c3}{3/3}
\psfrag{n6c3}{6/3}
\psfrag{n6c6}{6/6}
\includegraphics[width=.48\textwidth]{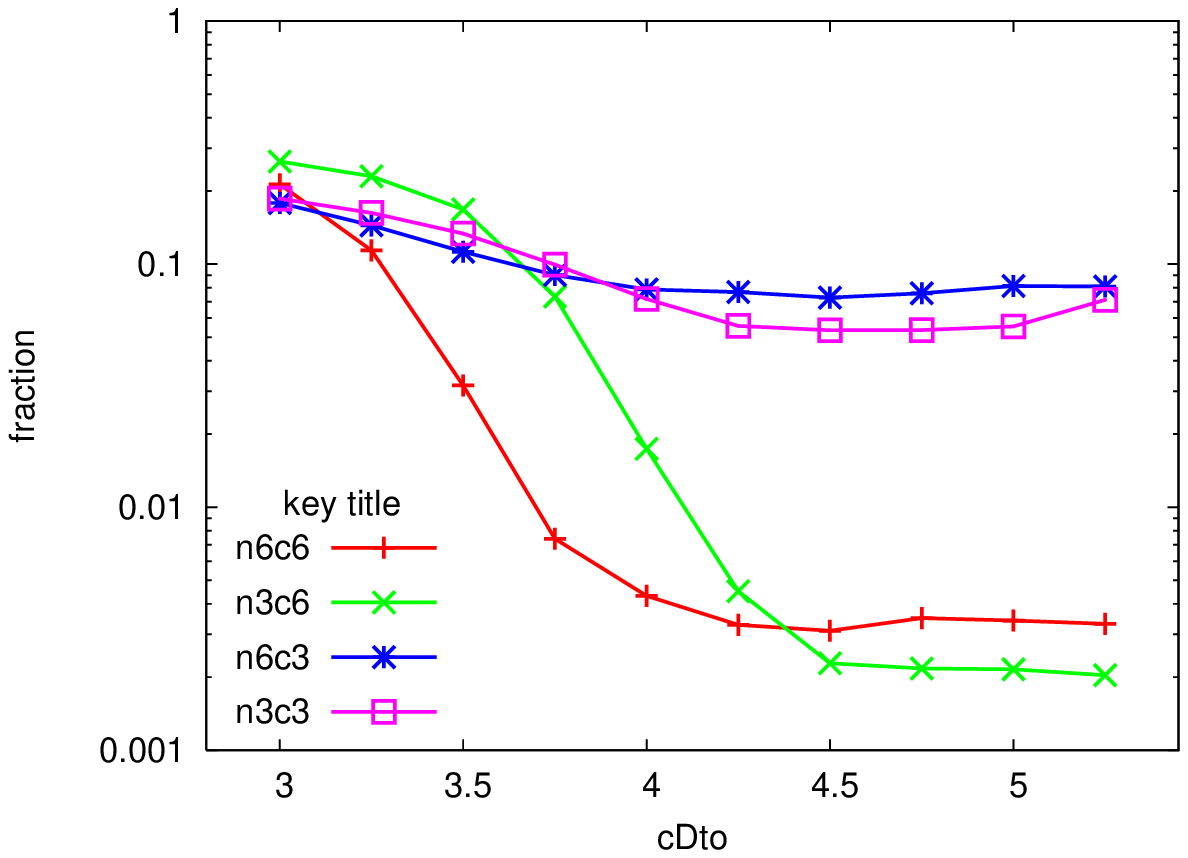}
%}
%\hfill
%\subfigure[\, varying $c_{\mathrm{E,to}}$]{
\psfrag{cEto}{$c_{\mathrm{E,to}}\;[\mu\text{M}]$}
\psfrag{fraction}[c]{fraction of skipped beats}
\psfrag{key title}{$n_{\mathrm{nuc}}/n_{\mathrm{cap}}$}
\psfrag{n3c6}{3/6}
\psfrag{n3c3}{3/3}
\psfrag{n6c3}{6/3}
\psfrag{n6c6}{6/6}
\includegraphics[width=.48\textwidth]{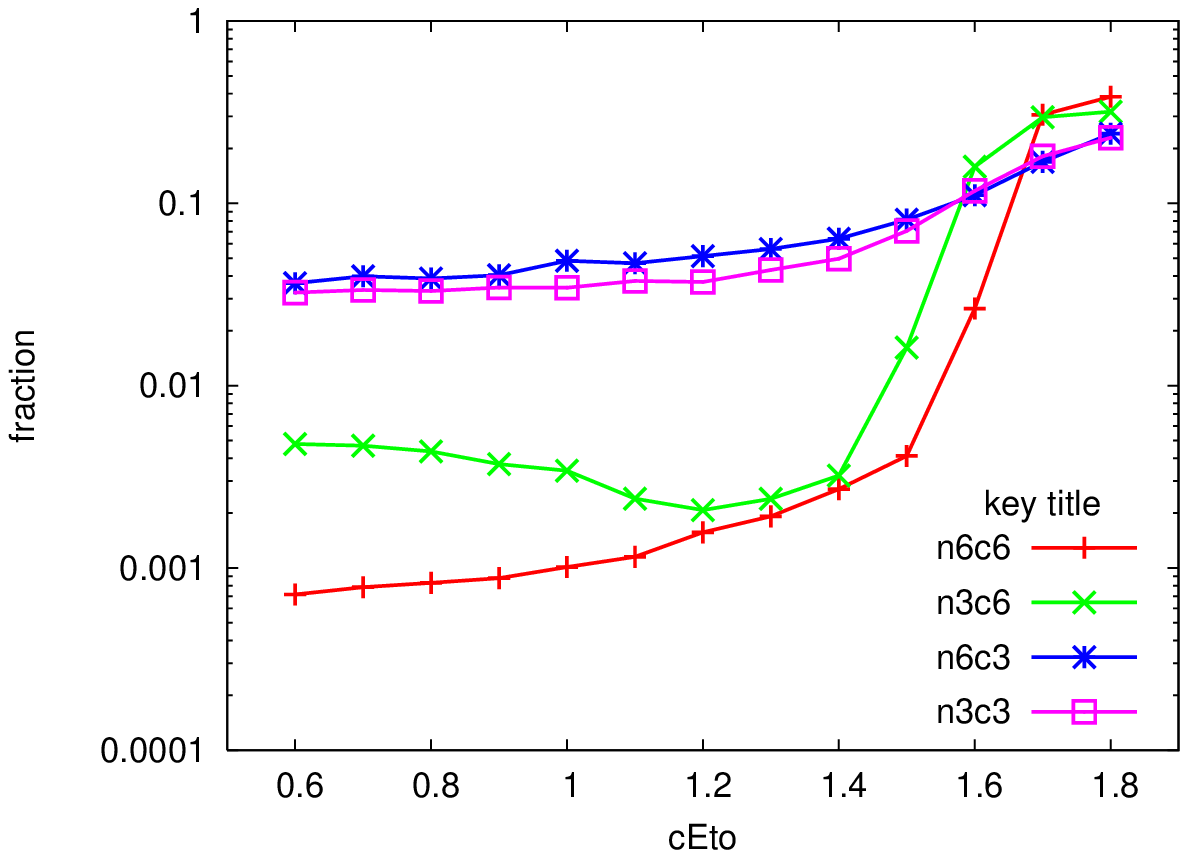}
%}
\caption{Numerical result for the regularity of the oscillations. Plotted is the fraction of skipped beats (see text) depending on both total concentrations of the Min proteins and cooperativities used for nucleation and capping, respectively. In the upper panel, $c_{\mathrm{E,to}}=1.5\,\mathrm{\mu M}$, in the lower $c_{\mathrm{D,to}}=4\,\mathrm{\mu M}$. Simulations were run such that the total number of cappings are in the range of $10^5$ for the low values and $10^3 - 10^5$ for the high values. The errorbars are smaller than the pointsize and the lines are guides to the eye.}
\label{fig:beat_skipping}
\end{figure}

Stochastic switching (nucleation and capping) can also lead to occasional skipping of beats, another property that is easily quantified experimentally. Regular oscillations are characterized by the asynchronous and alternating growth and disassembly of a polymer at each of the two poles of the cell. A skipped beat is defined as a deviation from regular oscillations in which alternation fails and a polymer reappears on the same pole before the appearance of a polymer on the opposite pole. As a criterion for our numerical analysis, we count the number of cases where two consecutive cappings of MinD polymers occur on the same side of the cell. The fraction of these cases with respect to all cappings is shown in Fig.~\ref{fig:beat_skipping}. The figure shows numerically-obtained data for different cooperativities of nucleation and capping.

As a skipped beat is one way of deviating from regular oscillations, the results in Fig.~\ref{fig:beat_skipping} recapitulate the observations described earlier in the context of Figs.~\ref{fig:distributions_numerically}, \ref{fig:T_hist}, and~\ref{fig:mean_T}. The high fraction of irregular capping events for low total MinD concentrations as well as for high total MinE concentration correspond to the second peak in the probability distributions of periods in Fig.~\ref{fig:T_hist} at short $T$. If the capping cooperativity is relatively low, there will always be a significant fraction of skipped beats, which corresponds to the non-zero tails of the distributions at short lengths for $n_{\mathrm{cap}}=3$ in Fig.~\ref{fig:distributions_numerically}. For our standard parameter set $n_{\mathrm{nuc}}=6, n_{\mathrm{cap}}=6, c_{\mathrm{D,to}}=4\,\mathrm{\mu M}, c_{\mathrm{E,to}}=1.5\,\mathrm{\mu M}$ and for a range of concentrations around it, the fraction of skipped beats is well below the 1\% figure. The strong deviation from our analytical result when $c_{\mathrm{E,to}}$ is very small (Fig.~\ref{fig:distr_sim_int_c_lowE}) is a violation of the regular oscillation pattern that does not lead to skipped beats, i.e. even in this extreme case the alternating appearance of a polymer on the left and the right side is still conserved.

%{\it Why is 3/6 smaller than 6/6 in (a)? Why is 3/3 smaller than 6/3?}

%%%%%%%%%%%%%%%%%%%%%%%%%%%%%%%%%%%%%%%%%%%%%%%%%%%%%%%%%%%%%%%%%%%%%%%%%%%%%

\subsection{Bistability and stochastic transitions}
\label{ssec:transitions}

\begin{figure*}
\psfrag{t [s]}[c]{$t\;[\mathrm{s}]$}
\psfrag{x [mum]}{$x\;[\mathrm{\mu m}]$}
\centering
\subfigure[\, $n_{\mathrm{cap}}=6$, $c_{\mathrm{E,to}}=1.7$, $k_{\mathrm{cap}}=0.15$]{
\label{fig:episodes1:17}
\includegraphics[width=.21\textwidth]{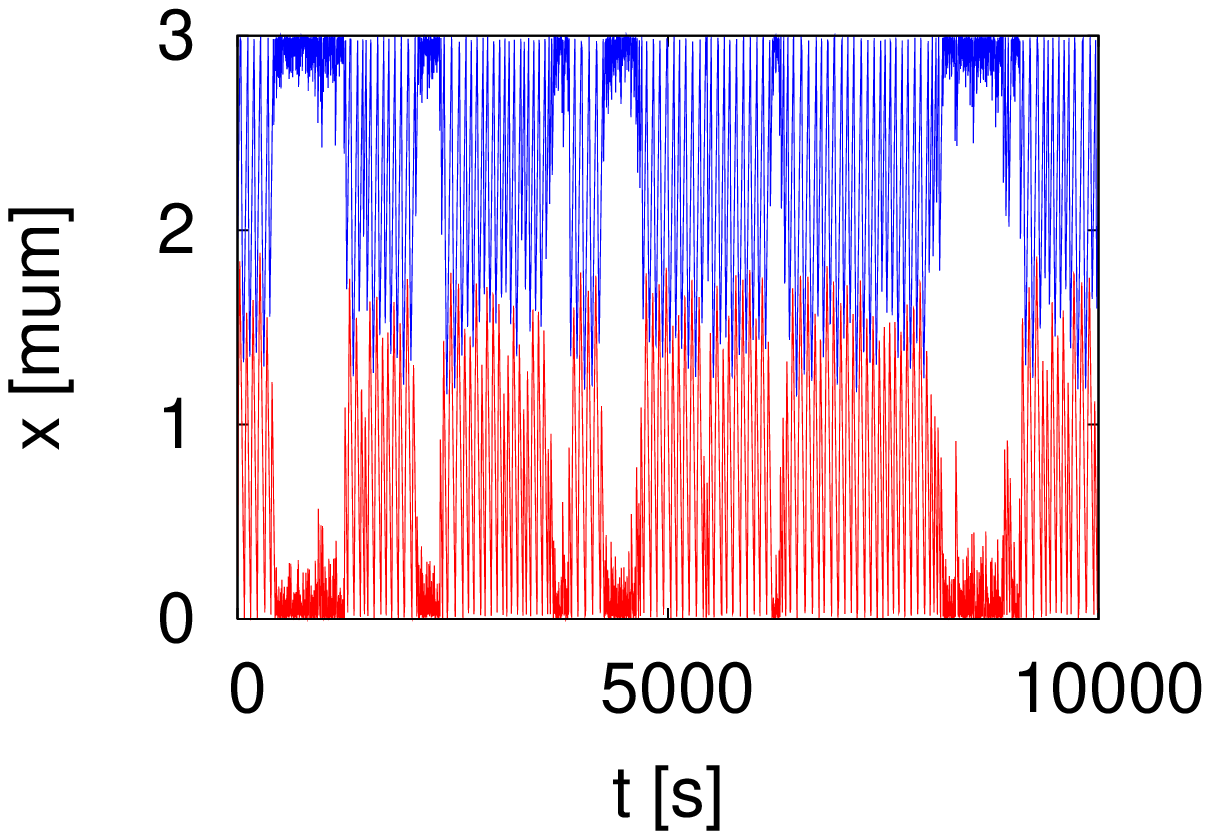}}
\subfigure[\, $n_{\mathrm{cap}}=6$, $c_{\mathrm{E,to}}=1.5$, $k_{\mathrm{cap}}=0.5$]{
\label{fig:episodes1:15}
\includegraphics[width=.21\textwidth]{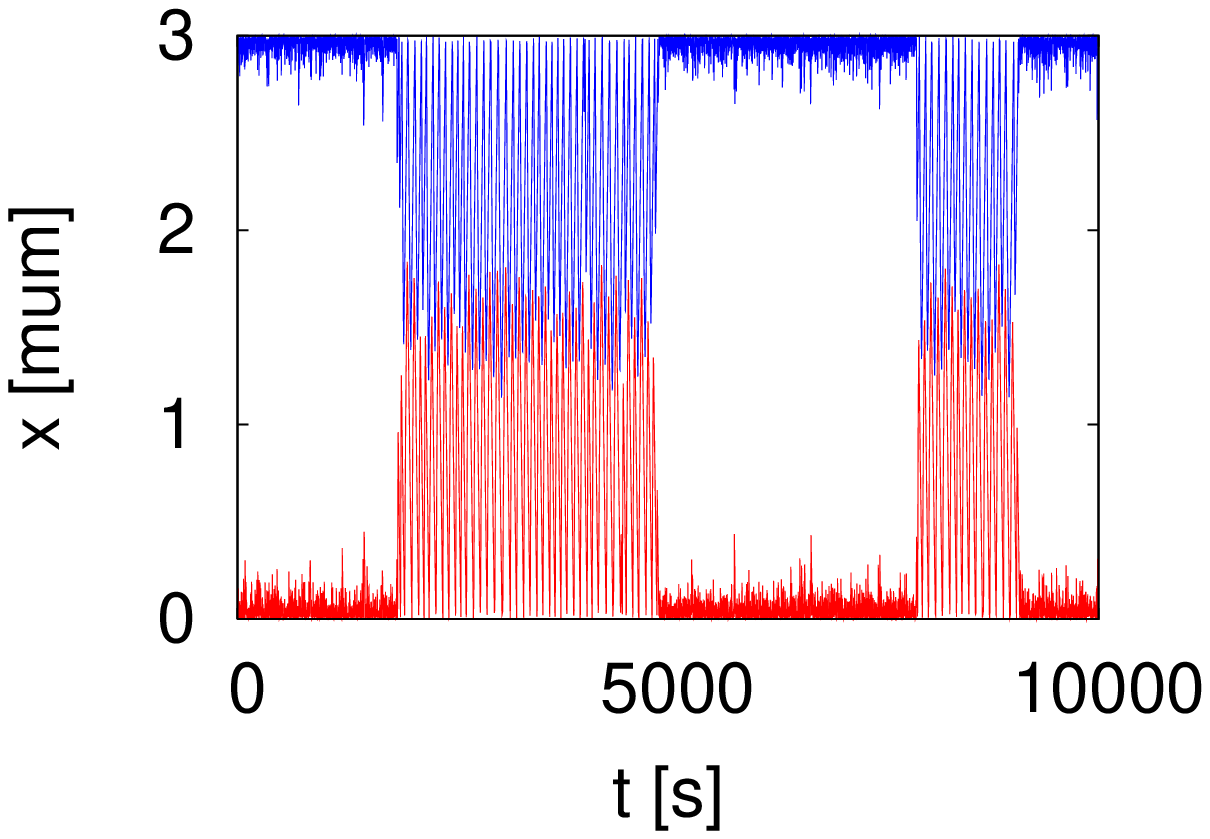}}
\subfigure[\, $n_{\mathrm{cap}}=3$, $c_{\mathrm{E,to}}=1.5$, $k_{\mathrm{cap}}=0.4$]{
\label{fig:episodes1:3}
\includegraphics[width=.21\textwidth]{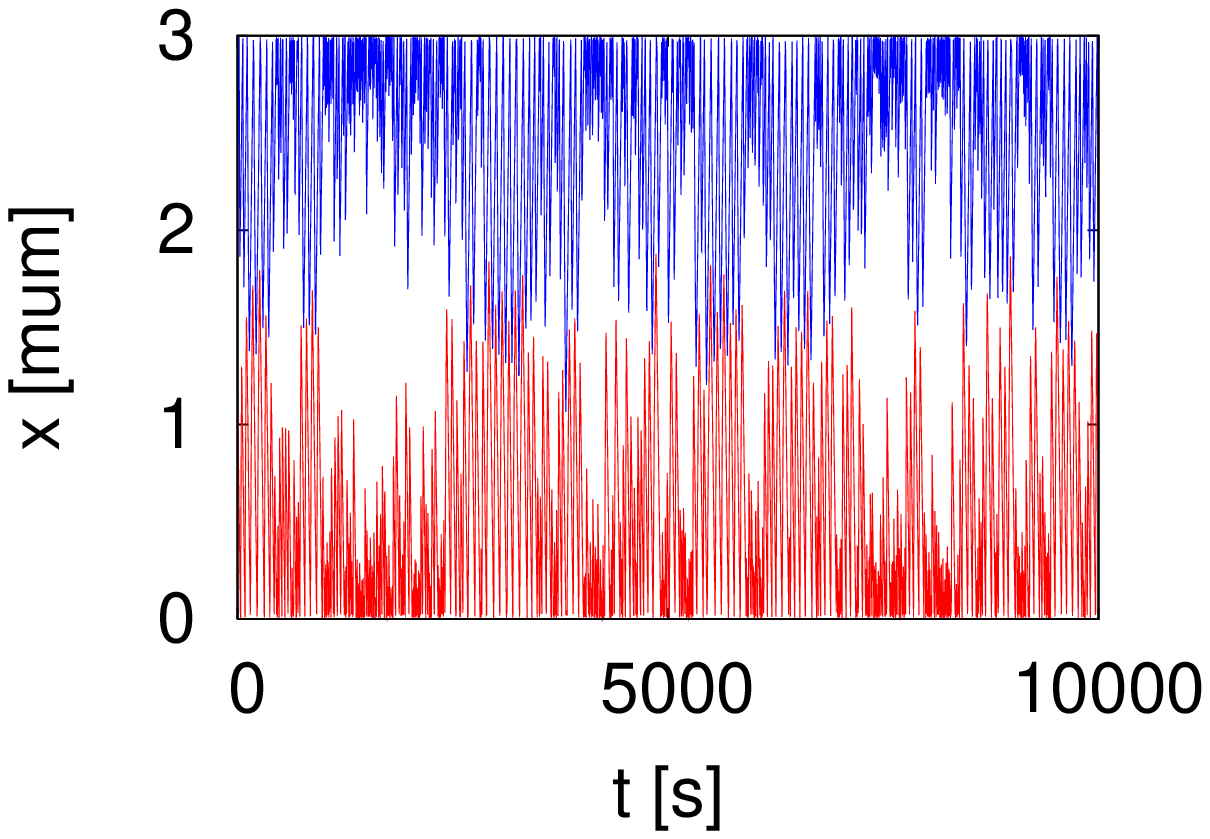}}
\subfigure[]{
\psfrag{one}[cb]{Fig.~\ref{fig:distr_sim_int_c_standard}}
\psfrag{17}{10(a)}
\psfrag{15}[cb]{\, 10(b)}
\psfrag{l1}[c]{$l^c_i\;[\mathrm{\mu m}]$}
\psfrag{l2}[c]{$l^c_{i+1}\;[\mathrm{\mu m}]$}
\label{fig:episodes1:map}
\includegraphics[width=.27\textwidth]{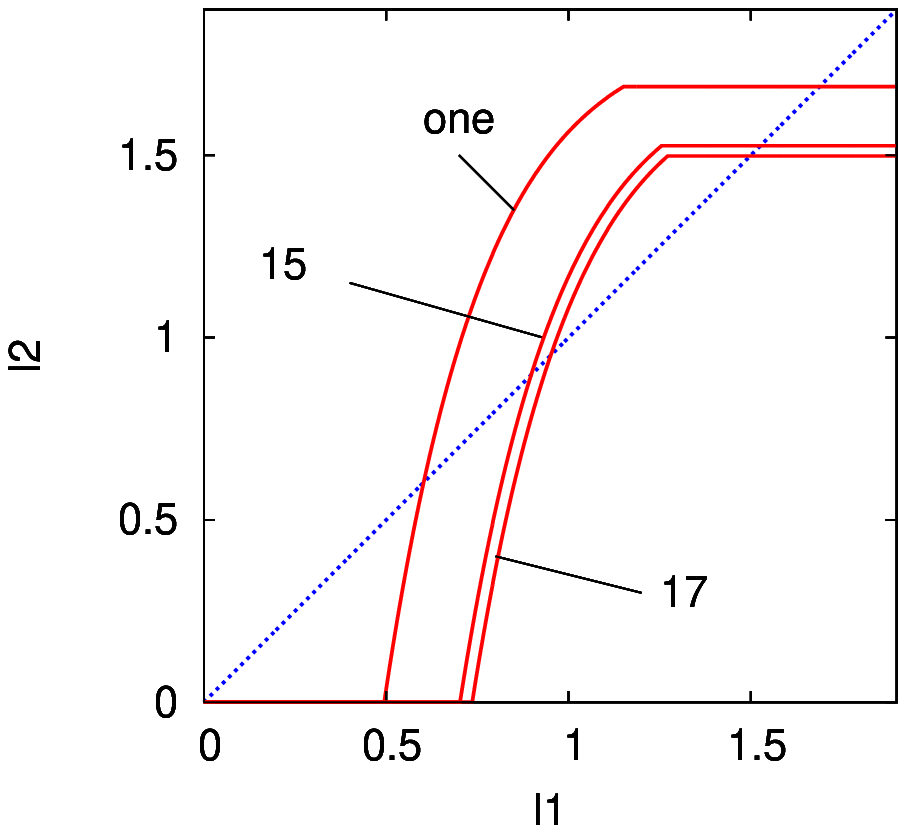}}
\caption{Episodes of oscillatory dynamics alternating with the cytosolic state. Plotted is the position (along the cell long axis $x$) of the two D-polymer tips over time. Note that, typically, the duration of one period in the oscillatory state is around $T\sim 100\;\mathrm{s}$. The map shown in (d) is used for a qualitative explanation in the text. Note, that contrary to our discussion in the text, the time series presented in (a)--(c) were obtained from the full stochastic model (nucleation and capping stochastic).}
\label{fig:episodes1}
\end{figure*}

As derived in Subsec.~\ref{ssec:det}, the deterministic system is bistable for a large range of parameters. Under certain conditions, stochastic nucleation and capping of polymers can lead to transitions between the two stable states: oscillatory episodes lasting for tens of periods and the cytosolic state. Fig.~\ref{fig:episodes1} shows examples of this behaviour for three different parameter sets. A clear distinction between regular oscillations and the cytosolic state is only possible for high cooperativity. This behaviour can be explained qualitatively using the deterministic map described in Subsec.~\ref{ssec:det}. 

During regular oscillations, a D-polymer is most likely capped at a length close to the peak of the respective conditional probability distribution (Fig.~\ref{fig:Pcapt1cap} in App.~\ref{app:cap_distr}). Due to the stochasticity of the capping process, there is a small probability of the D-polymer being capped before it reaches normal extension\footnote{If the capping occurs while the E-polymer on the opposite side still has its steady state length $l_{\mathrm{E,ss}}$, our model predicts a second E-ring of length zero. According to our model equations, the E-polymer could nucleate (stochastically) but the deterministic growth equation does not support an elongation because $c_{\mathrm{E}}$ is at the critical concentration for elongation due to the presence of the other E-ring. A more detailed model would incorporate stochastic effects in the growth and shrinking of the E- (and DE-) polymer.}. For high cooperativities, the deterministic map in Fig.~\ref{fig:map} retains some validity and one can think of the capping events in the stochastic model as being a blurred version of the deterministic map. Rare early and late capping events correspond to events at the edge of the blurred region about each fixed point and can result in a transition from relatively stable oscillations to a state that is mostly cytosolic and back again. Since the events allowing for such transitions lie outside the regular oscillation pattern (Subsec.~\ref{ssec:model_summary}), the analytical approach in App.~\ref{app:map_prob} does not provide the right tools to describe this effect. We restrict ourselves to a qualitative discussion of a limiting case.

In order to get a better understanding of the stochastic transitions we consider the special case of deterministic nucleation of D-polymers and stochastic capping\footnote{In the opposite case (stochastic nucleation and deterministic capping), the transitions are not observed. In this case, if the regular oscillation pattern is violated and a D-polymer nucleates very late, $c_{\mathrm{E}}>c_{\mathrm{E,th}}$, and the D-polymer as well as all following D-polymers will be capped right away.}. For high values of $n_{\mathrm{cap}}$, the map in Fig.~\ref{fig:map} can be used as a rough guide for the stochastic transitions. In App.~\ref{app:underlying_map}, we adjust this map to the specific case considered here.

In Fig.~\ref{fig:episodes1:map} we show three maps obtained as in App.~\ref{app:underlying_map}: One representing our standard parameter set $c_{\mathrm{E,to}}=1.5\;\mathrm{\mu M}$, $k_{\mathrm{cap}}=0.15\,\mathrm{s^{-1}\mu M^{-6}}$ (cf. Fig.~\ref{fig:distr_sim_int_c_standard}) and the two parameter sets used in Fig.~\ref{fig:episodes1:17} and~\ref{fig:episodes1:15} (with $c_{\mathrm{D,th}}=2.5\,\mathrm{\mu M}$). One can think of the stochastic system as producing a cloud of points in this map around the stable steady states. A transition occurs if an extreme capping event takes the system from a neighbourhood of the pre-transition fixed point over the unstable steady state into a neighbourhood of the other fixed point. For our standard parameter set, due to the large distance between stable and unstable fixed points, this is unlikely to happen. The system will stay in the vicinity of one of the fixed points with probability close to one. When the total E-concentration is increased (Fig.~\ref{fig:episodes1:17}), the map is shifted to the right and the distance an extreme capping event has to cover in order for a transition to occur, decreases. The same is valid for an increase in the capping parameter $k_{\mathrm{cap}}$ as observed for Fig.~\ref{fig:episodes1:15}. The map for this case is slightly shifted to the left compared to the former case which leads to a reduced transition rate. The main parameter controlling the extension of the cloud of points is the capping cooperativity $n_{\mathrm{cap}}$. When it decreases, the cloud becomes larger, rendering the map interpretation less meaningful (cf. Fig.~\ref{fig:distributions_numerically}). More transitions occur and the states are not as well defined anymore (Fig.~\ref{fig:episodes1:3}).

%%%%%%%%%%%%%%%%%%%%%%%%%%%%%%%%%%%%%%%%%%%%%%%%%%%%%%%%%%%%%%%%%%%%%%%%%%%%%

\section{Discussion}

The mechanisms underlying the Min oscillations in E.coli are still subject to much debate among modellers since none of the models proposed so far capture all the properties and all the phenotypes displayed in wildtype and mutant cells (see~\cite{Cytrynbaum2007} for a discussion). The most obvious qualitative features of the observed dynamics are explained in many models which means that being able to distinguish between these models will rely on careful quantitative characterization of both the models and the experimental data. The results presented here are a step in this direction. 

We generalized a recently published model to allow for more realistic comparison with data by introducing and analyzing stochasticity. From a mathematical viewpoint, this model provides an example of a stochastic hybrid dynamic system, whose solution can be found analytically in the case of regular oscillations. Probability distributions for easily measured quantities are provided, making the model testable. Comparing our results to experimental data already available in the literature, the main findings of this paper are as follows. 

High cooperativity in the nucleation of the Min-polymers (at least a power of 3 for the MinD and 6 for MinE) are important. Lower cooperativities allow too many events that are inconsistent with observations. Reported Hill coefficients for the nucleation of MinD on a lipid membrane are around 2~\cite{Mileykovskaya2003}.  

The model is relatively stable to changes in MinD concentration. MinD concentrations greater than $3.5 \,\mathrm{\mu M}$ ensure regular oscillations. MinE concentration, on the other hand, is more finely constrained and must fall between 1.2 and $1.6 \,\mathrm{\mu M}$. 

Our results offer an explanation for the variability in the presence of oscillations observed in experiments. A significant fraction of cells usually do not show oscillations in experiments or go from an oscillatory state into a cytosolic state during the course of observation (e.g.~\cite{Raskin1999}). Our model suggests an explanation for this in the form of bistability of an oscillating state and a purely cytosolic state. The model of Fange and Elf~\cite{Fange2006} also reports bistability, whereas the models of Howard et al.~\cite{Howard2001} and Meacci et al.~\cite{Meacci2005} demonstrate qualitatively distinct solutions (cytosolic or oscillating) as a function of parameters but not bistability. Transitions between the two states will be a challenge to observe experimentally. Evidence for a transition in each direction would be necessary, requiring long observation times. Photobleaching in the fluorescence microscopy studies limits the latter and other factors might influence the existence and quality of the observed oscillations (see, e.g.~\cite{Raskin1999}). 

In a recent experimental study, Downing et al.~\cite{Benjamin} actively controlled the MinD dynamics in E.coli by changing cationic concentrations in the surrounding medium. They were able to stop, distort and restart the oscillations. Based on the bistability and stochastic transitions described here, we speculate that the ion concentration has an influence on the assembly dynamics of the MinE-polymer. For example, if increased ion concentration increases $k_{\mathrm{cap}}$, the system might be driven into a regime where a stochastic transition into the cytosolic state becomes more likely. On the other hand, decreasing $k_{\mathrm{cap}}$ can lead to a freeze or or halt of the oscillations with most of the MinD localised on one side of the cell.

Most other models~\cite{Pavin2006,Meacci2005,Tostevin2006} produce a similar dependence of the oscillation period on the total Min concentrations and a few~\cite{Fange2006,Kerr2006} report on the variability of the period or, as in our case, the full probability distribution~\cite{Tostevin2006}.

The biggest difference between the reaction-diffusion-type models and the polymer model described here is that in the polymer model the spatial pattern of the Min oscillations is prescribed in the form of possible nucleation sites of MinD in the membrane. This assumption becomes most important when trying to model the striped Min phenotypes observed in filamentous cells~\cite{Raskin1999}. There is an ongoing debate in the experimental literature about the existence and possible nature of sites to which MinD binds preferentially. Recent work~\cite{Mileykovskaya2009} has shown that specific anionic phospholipids colocalise with MinD at the poles and septum. An important as yet unanswered question is which of these two determines the localisation of the other or if there is some other factor upstream of both.

%{\it comparison to other models:  possible difference to Fange: Fig. 4c/d -- lower MinD in polymer model leads to shorter D-polymers $\rightarrow$ maybe $<.5$ cell length $\rightarrow$ no MinD midcell at any time.}

% Even within this limit, however, the analytical solution can only be expressed in terms of non-trivial integrals that must be solved numerically. We nevertheless think the derivation presented in this article might be instructive to researches in this field.

%In summary, we hope that this article -- together with~\cite{Cytrynbaum2007} and an analysis of the polymer model in filamentous mutants that will be presented elsewhere -- offers enough results to discriminate the polymer model against other prevailing models. Since our model agrees with most published experimental results, more quantitative experimental data is needed to achieve this point. 

%%%%%%%%%%%%%%%%%%%%%%%%%%%%%%%%%%%%%%%%%%%%%%%%%%%%%%%%%%%%%%%%%%%%%%%%%%%%%

\appendix

\renewcommand{\theequation}{\Alph{section}.\arabic{equation}}

\section{The deterministic version as a one-dimensional map}
\label{app:map}
\setcounter{equation}{0}

In its simplest version (deterministic switching and fast E-ring formation, cf. beginning of Sec.~\ref{sec:results}), our model allows for a fully analytical solution in terms of a discrete map. We choose to display the maximal length of the MinD polymer as a function of the temporally previous maximal length on the other side of the cell: $l^{c}_{i+1} = f(l^c_i)$ (cf. Fig.~\ref{fig:Ts_and_ls}). 

To derive the map, we consider the following case: At time $t=0$, the DE-polymer on the left side ($l$) is decaying ($S^{\mathrm{D}}_l(0)=0$) and the D-polymer on the right side ($r$) is growing. Without loss of generality, we choose $t=0$ such that the DE-polymer on the left just reached the length $l_l(0)=\frac{\gamma V}{\lambda}(c_{\mathrm{E,to}}-c_{\mathrm{E,th}})$, which causes the cytosolic E-concentration to grow above $c_{\mathrm{E,th}}$. This causes the D-polymer on the right side to be capped: $t^c_{r}=0$ and therefore its state to be switched: $S^{\mathrm{D}}_r: 1\rightarrow 0$. We assume that the D-polymer on the right had a length $l_r(0)=l^c_r$ at the time point of capping. For a short period of time, now both polymers are decaying: $l_{l/r}(t) = l_{l/r}(0)-\gamma k_{\mathrm{off}}t$. At time $t^d_l=\frac{1}{\gamma k_{\mathrm{off}}}l_l(0)$, the left nucleation site becomes free for a new nucleation.

We now have to distinguish three different cases:

\begin{enumerate}
\item The newly nucleated polymer on the left side gets capped immediately if $c_{\mathrm{E}}(t^{d}_l)>c_{\mathrm{E,th}}$. This is the case if
\begin{equation}
l^c_r < 2\frac{\gamma V}{\lambda}(c_{\mathrm{E,to}}-c_{\mathrm{E,th}}) .
\label{eq:map_1}
\end{equation}
Under this condition, the cytosolic MinE concentration will never drop below $c_{\mathrm{E,th}}$. Any nucleating MinD polymer will therefore be capped immediately, which evolves into the cytosolic state (i.e. no polymer: $l^c_l=0$).

\item For slightly larger $l^c_r$, nucleation on the left side will happen right after the first polymer (left) completely disassembled ($l_l=0$; $t^n_l=t^d_l$), as long as $c_{\mathrm{D}}(t^d_l) >c_{\mathrm{D,th}}$. Using $t^c_l=\frac{1}{\gamma k_{\mathrm{off}}}[l^c_r-\frac{\gamma V}{\lambda}(c_{\mathrm{E,to}}-c_{\mathrm{E,th}})]$ in Eq.~\ref{eq:lD_t} leads to an expression for the length of the D-polymer at the next capping:
\begin{align}
l^c_l = & \frac{\gamma V}{\lambda} \left[ c_{\mathrm{D,to}} - \frac{k_{\mathrm{off}}}{k^{\mathrm{D}}_{\mathrm{on}}}-c_{\mathrm{E,to}}+c_{\mathrm{E,th}} \right. \nonumber \\
& \left. - \left( c_{\mathrm{D,to}} - \frac{k_{\mathrm{off}}}{k^{\mathrm{D}}_{\mathrm{on}}} + c_{\mathrm{E,to}} - c_{\mathrm{E,th}} - \frac{\lambda}{\gamma V} l^c_r\right) \right. \nonumber \\ 
& \hspace{-1cm} \left. \cdot \exp\left( -\frac{\lambda k^{\mathrm{D}}_{\mathrm{on}}}{\gamma k_{\mathrm{off}} V}\left( l^c_r - 2\frac{\gamma V}{\lambda}(c_{\mathrm{E,to}}-c_{\mathrm{E,th}})\right)\right)\right] .
\label{eq:map_2}
\end{align}

\item If $l^c_r$ is even bigger ($l^c_r\ge \frac{\gamma V}{\lambda}(c_{\mathrm{D,th}}+c_{\mathrm{E,to}}-c_{\mathrm{E,th}})$), such that $c_{\mathrm{D}}(t^d_l) <c_{\mathrm{D,th}}$, the new polymer on the left will nucleate at $t^n_l=\frac{1}{\gamma k_{\mathrm{off}}}[l^c_r-\frac{\gamma V}{\lambda}(c_{\mathrm{D,to}}-c_{\mathrm{D,th}})]$. Its capping will then happen at the stable amplitude:
\begin{align}
l^c_l = & \frac{\gamma V}{\lambda} \left[ c_{\mathrm{D,to}} - \frac{k_{\mathrm{off}}}{k^{\mathrm{D}}_{\mathrm{on}}}-c_{\mathrm{E,to}}+c_{\mathrm{E,th}} \right. \nonumber \\
 & \left. - \left( c_{\mathrm{D,th}} - \frac{k_{\mathrm{off}}}{k^{\mathrm{D}}_{\mathrm{on}}}\right) \right. \nonumber \\ 
& \left. \cdot \exp\left( -\frac{k^{\mathrm{D}}_{\mathrm{on}}}{k_{\mathrm{off}}}\left( c_{\mathrm{D,to}}-c_{\mathrm{D,th}}-c_{\mathrm{E,to}}+c_{\mathrm{E,th}}\right)\right)\right] .
\label{eq:map_3}
\end{align}

This is the amplitude of the stable oscillation and all consecutive cappings will happen at this amplitude, too.

\end{enumerate}

The piecewise map derived above is displayed for various MinD and MinE concentrations in Fig.~\ref{fig:map}.

%%%%%%%%%%%%%%%%%%%%%%%%%%%%%%%%%%%%%%%%%%%%%%%%%%%%%%%%%%%%%%%%%%%%%%%%%%%%%

\section{Conditional probability distribution function for nucleation times (regular oscillation)}
\label{app:nuc_distr}
\setcounter{equation}{0}

In the stochastic version of our model, the probability distribution function of the times one side of the cell is free of polymer can be derived analytically if the system is oscillating regularly (see Subsec.~\ref{ssec:model_summary}). This distribution function is conditional on the length $l^d$ of the polymer present on the opposite side of the cell at the time point where the nucleation site becomes polymer free ($t^d$ in Fig.~\ref{fig:Ts_and_ls}). As described in Subsec.~\ref{ssec:model_switching}, we assume the instantaneous probability for nucleation on a free nucleation site at time $t$ to be $\lambda_{\mathrm{nuc}}(t)=k_{\mathrm{nuc}}c_{\mathrm{D}}^{n_{\mathrm{nuc}}}(t)$.

The probability for nucleation between times $t$ and $t+\dd t$ is the probability of no nucleation until $t$ and then nucleating in $t..(t+\dd t)$: $P_{\mathrm{nuc}}(t)\dd t = (1-P(t^n<t))\lambda_{\mathrm{nuc}}(t)\dd t$. $P(t^n<t)$ is the cumulative distribution function. Without loss of generality, we assume $t^d_l=0$ (i.e., the polymer on the left side just decayed completely at time $t=0$) and obtain
\begin{equation}
P_{\mathrm{nuc}}(t) = \lambda_{\mathrm{nuc}}(t)\exp\left(-\int_0^t \lambda_{\mathrm{nuc}}(t')\dd t'\right).
\label{eq:Pnuct1}
\end{equation}

The cytosolic D-concentration follows $c_{\mathrm{D}}(t)=c_{\mathrm{D,to}}-\frac{\lambda}{\gamma V}(l_r^d-\gamma k_{\mathrm{off}}t)$ for $t\le t^d_r$ (on the other side). For $t\ge t^d_r$, there would only be cytosolic MinD ($c_{\mathrm{D}}(t)=c_{\mathrm{D,to}}$). We do not consider this case, since it is outside the regular oscillation pattern.

Putting $c_{\mathrm{D}}(t)$ from above into Eq.~\ref{eq:Pnuct1}, one obtains
\begin{align}
& P_{\mathrm{nuc}}  (T^f|l^d) = k_{\mathrm{nuc}} \left( c_{\mathrm{D,to}} - \frac{\lambda}{\gamma V}l^d + \frac{\lambda}{V}k_{\mathrm{off}}T^f\right)^{n_{\mathrm{nuc}}} \nonumber \\
 & \; \cdot \exp\left[ -\frac{1}{n+1}\frac{k_{\mathrm{nuc}}V}{\lambda k_{\mathrm{off}}}\left( \left( c_{\mathrm{D,to}} - \frac{\lambda}{\gamma V}l^d + \frac{\lambda}{V}k_{\mathrm{off}}T^f\right)^{n_{\mathrm{nuc}}+1} \right.\right. \nonumber \\
 & \qquad \left.\left. - \left( c_{\mathrm{D,to}} - \frac{\lambda}{\gamma V}l^d\right)^{n_{\mathrm{nuc}}+1}\right)\right] .
\label{eq:nuc_distr}
\end{align}

Fig.~\ref{fig:Pcapt1} shows typical shapes of this probability distribution function (for a given $l^d$) for high cooperativity in nucleation and different concentrations of MinD. From Eq.~\ref{eq:nuc_distr}, it is obvious that the dependence of the probability distribution on $l^d$ is (up to a scaling factor) the same as on $c_{\mathrm{D,to}}$.

%%%%%%%%%%%%%%%%%%%%%%%%%%%%%%%%%%%%%%%%%%%%%%%%%%%%%%%%%%%%%%%%%%%%%%%%%%%%%

\section{Conditional probability distribution function for capping times (regular oscillation)}
\label{app:cap_distr}
\setcounter{equation}{0}

Equivalently to the preceding appendix, one can derive an expression for the conditional probability distribution function for the time of growth of a polymer. In Subsec.~\ref{ssec:model_switching} we introduced the instantaneous probability for capping of a growing polymer at a time $t$: $\lambda(t) = k_{\mathrm{cap}}c_{\mathrm{E}}^{n_{\mathrm{cap}}}(t)$. 

We only consider the approximative case of fast E-ring formation (see Sec.~\ref{sec:results}). The MinDE-polymer on the right side is assumed to be decaying and has length $l^n_r$ at time $t=0$ (cf. Fig.~\ref{fig:Ts_and_ls}). Without loss of generality we assume that the D-polymer on the left side nucleates and starts growing at time $t^n_l=0$. We now have to distinguish three cases during the decay of the polymer on the right side ($t^{d,\mathrm{E}}$ is the time when the growing tip of the E-polymer reaches the cell wall):
\begin{enumerate}
\item First, the concentration of MinE monomers in the cytosol is constant ($l^{\mathrm{E}}_r=l_{\mathrm{E,ss}} \; \Rightarrow \; c_{\mathrm{E}} = c_{\mathrm{E},0} \equiv \frac{k_{\mathrm{off}}}{k^{\mathrm{E}}_{\mathrm{on}}}$).
\item For $t^{d,\mathrm{E}}_r < t < t^d_r$, $c_{\mathrm{E}}$ grows linearly: $c_{\mathrm{E}}(t)=c_{\mathrm{E},0}+\frac{\lambda}{V}k_{\mathrm{off}}(t-t^{d,\mathrm{E}}_r)$.
\item After $t^d_r$, $c_{\mathrm{E}}$ is constant again: $c_{\mathrm{E}}=c_{\mathrm{E,to}}$.
\end{enumerate}
Putting these into the equivalent of Eq.~\ref{eq:Pnuct1} gives the conditional probability distribution function for the time, at which the growing D-polymer on the left side gets capped and switches states (see Eq.~\ref{eq:Pcapt_final3} -- we generalise to $T^c$ and drop the side-dependence; $c_{\mathrm{E},0}=\frac{k_{\mathrm{off}}}{k_{\mathrm{on}}^{\mathrm{E}}}$; $A=\frac{\lambda}{V}k_{\mathrm{off}}$).
\begin{figure*}
\begin{equation}
P_{\mathrm{cap}}(T^c|l^n) = \left\{ \begin{array}{ll} k_{\mathrm{cap}}c_{\mathrm{E},0}^{n_{\mathrm{cap}}}\exp\left( -k_{\mathrm{cap}}c_{\mathrm{E},0}^{n_{\mathrm{cap}}} T^c \right) &  \quad 0\le T^c < t^{d,\mathrm{E}} \\[2mm]
k_{\mathrm{cap}}(c_{\mathrm{E},0}+A(T^c-t^{d,\mathrm{E}}))^{n_{\mathrm{cap}}} &  \\
  \quad \cdot \exp\left(-k_{\mathrm{cap}}c_{\mathrm{E},0}^{n_{\mathrm{cap}}} t^{d,\mathrm{E}}-\frac{k_{\mathrm{cap}}}{A(n_{\mathrm{cap}}+1)}\left[ (c_{\mathrm{E},0}+A(T^c-t^{d,\mathrm{E}}))^{n_{\mathrm{cap}}+1} - c_{\mathrm{E},0}^{n_{\mathrm{cap}}+1}\right]\right) &  \quad t^{d,\mathrm{E}}\le T^c < t^d \\[2mm]
k_{\mathrm{cap}}c_{\mathrm{E,to}}^{n_{\mathrm{cap}}} \exp\left( -k_{\mathrm{cap}}c_{\mathrm{E},0}^{n_{\mathrm{cap}}} t^{d,\mathrm{E}}-\frac{k_{\mathrm{cap}}}{A(n_{\mathrm{cap}}+1)} \right. &  \\
 \left. \quad \cdot \left[(c_{\mathrm{E},0}+A(t^d-t^{d,\mathrm{E}}))^{n_{\mathrm{cap}}+1}-c_{\mathrm{E},0}^{n_{\mathrm{cap}}+1}\right] - k_{\mathrm{cap}}c_{\mathrm{E,to}}^{n_{\mathrm{cap}}}(T^c-t^d)\right)  &  \quad  T^c \ge t^d
\end{array}
\right.
\label{eq:Pcapt_final3}
\end{equation}
\end{figure*}
%with $c_{\mathrm{E},0}=\frac{k_{\mathrm{off}}}{k_{\mathrm{on}}^{\mathrm{E}}}$ and $A=\frac{\lambda}{V}k_{\mathrm{off}}$.
Here, $t^d=\frac{1}{\gamma k_{\mathrm{off}}}l^n$ and $t^{d,\mathrm{E}}=\frac{1}{\gamma k_{\mathrm{off}}}(l^n-l_{\mathrm{E,ss}})$ with $l_{\mathrm{E,ss}}=\frac{\gamma V}{\lambda}(c_{\mathrm{E,to}}-\frac{k_{\mathrm{off}}}{k_{\mathrm{on}}^{\mathrm{E}}})$. The condition on $l^n$ only enters through the dependencies in $t^{d,\mathrm{E}}$ and $t^d$. 

\begin{figure}
\centering
%\hfill
\subfigure[\, $P_{\mathrm{nuc}}(T^f|l^d)$,\; $l^d=1.5\;\mu\text{m}$]{
\psfrag{Tf}[c]{$T^f\;[\mathrm{s}]$}
\psfrag{p}{$p$}
\psfrag{key title}{$c_{\mathrm{D,to}}\;[\mathrm{\mu M}]$}
\psfrag{cDto=3.7}[c]{$c_{\mathrm{D,to}}=3.7$}
\psfrag{cDto=4}[c]{$c_{\mathrm{D,to}}=4$}
\psfrag{cDto=4.3}[c]{$c_{\mathrm{D,to}}=4.3$}
\includegraphics[width=.35\textwidth]{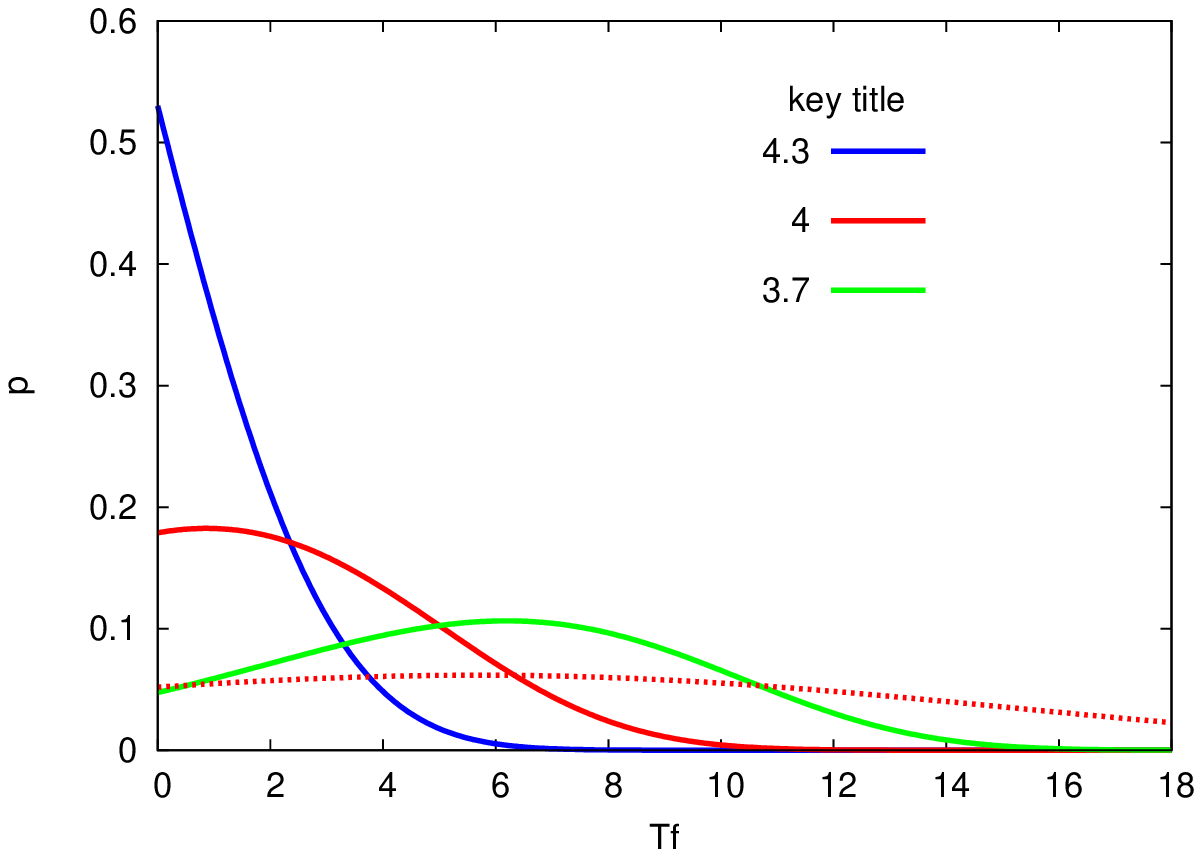}
\label{fig:Pcapt1nuc}
}
%\hfill
\subfigure[\, $P_{\mathrm{cap}}(T^c|l^n)$,\; $l^n=1.5\;\mu\text{m}$]{
\psfrag{Tg}[c]{$T^c\;[\mathrm{s}]$}
\psfrag{p}{$p$}
\psfrag{key title}{$c_{\mathrm{E,to}}\;[\mathrm{\mu M}]$}
\psfrag{cEto=1.3}[c]{$c_{\mathrm{E,to}}=1.3$}
\psfrag{cEto=1.5}[c]{$c_{\mathrm{E,to}}=1.5$}
\psfrag{cEto=1.7}[c]{$c_{\mathrm{E,to}}=1.7$}
\includegraphics[width=.35\textwidth]{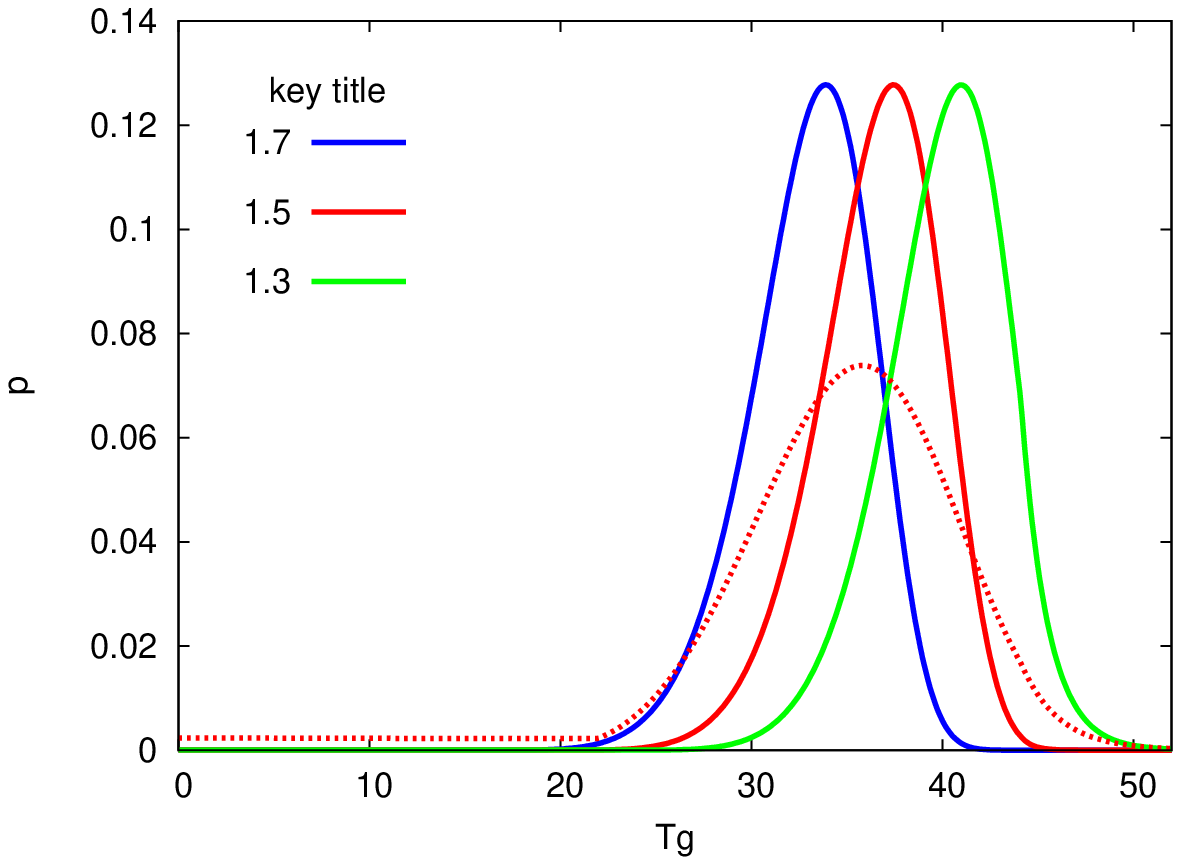}
\label{fig:Pcapt1cap}
}
%\hfill
\caption{Typical probability distributions for the time free of polymer $T^f$ (top) and the duration of growth $T^c$ (bottom), respectively. A number of curves are plotted for different total concentrations of MinD or MinE, respectively. The analytical expressions are given in Eq.~\ref{eq:nuc_distr} for the upper panel and Eq.~\ref{eq:Pcapt_final3} for the lower one. Standard parameters as of Tab.~\ref{tab:parameters} are used and the probability distributions are given for the conditions $l^d=1.5\,\mu\text{m}$ and $l^n=1.5\,\mu\text{m}$, respectively. The cooperativity for nucleation or capping is 6. For comparison, the dotted lines show the probability distribution functions for a reduced cooperativity $n_{\mathrm{nuc/cap}}=3$. Note the broader distributions for lower cooperativity as well as the noticeable tail towards shorter $T^c$ in the lower panel for the dotted line.}
\label{fig:Pcapt1}
\end{figure}

Fig.~\ref{fig:Pcapt1} shows typical probability distribution functions for capping (for a given $l^n$) for high capping cooperativity and different MinE concentrations. The dependence of the probability distribution on $l^n$ is (up to a scaling factor) the same as on $c_{\mathrm{E,to}}$ for the important first and second part of the non-smooth Eq.~\ref{eq:Pcapt_final3}.

%%%%%%%%%%%%%%%%%%%%%%%%%%%%%%%%%%%%%%%%%%%%%%%%%%%%%%%%%%%%%%%%%%%%%%%%%%%%%

\section{A map for the probability distribution}
\label{app:map_prob}
\setcounter{equation}{0}

Combining the two conditional probability distributions computed in Apps.~\ref{app:nuc_distr} and~\ref{app:cap_distr}, one can derive a map for the probability distribution of the amplitude during regular oscillations. Here, we derive this map in the following form: $P_n(l^n_{i})=F(P_n(l^n_{i-1}))$, i.e. the probability distribution of lengths at nucleation (on the opposite side) can be derived as a function of the distribution at the previous nucleation (on this side -- for notation see Fig.~\ref{fig:Ts_and_ls}).

We start with computing a relation between the two probability distributions $P_n(l^n_{i})$ and $P_d(l^d_{i})$:
\begin{equation}
P_n(l^n_{i}) = \int_{l^n_{i}}^{l_{\mathrm{max}}} p(l^n_{i}|l^d_{i}) P_d(l^d_{i}) \, \dd l^d_{i} .
\label{eq:distr_rel_1}
\end{equation}
The conditional probability distribution $p(l^n_{i}|l^d_{i})$ can be obtained from Eq.~\ref{eq:nuc_distr} by replacing $T^f=\frac{1}{\gamma k_{\mathrm{off}}}(l^d_{i}-l^n_{i})$ and multiplying with $\frac{1}{\gamma k_{\mathrm{off}}}$ (to ensure correct normalisation under the variable transformation). For the limits of the integration one has to consider that $l^n_{i}$ will always be smaller than $l^d_{i}$ (see Fig.~\ref{fig:Ts_and_ls}). The upper limit is the maximally possible length of a D-polymer (i.e. all MinD is bound in one polymer (Eq.~\ref{eq:lmax})).

With that, we obtain:
\begin{align}
P_n(l^n_{i}) = k_{\mathrm{nuc}} & \left( c_{\mathrm{D,to}}-\frac{\lambda}{\gamma V} l^n_i\right)^{n_{\mathrm{nuc}}} \exp \left[-\frac{1}{n_{\mathrm{nuc}}+1}\frac{Vk_{\mathrm{nuc}}}{\lambda k_{\mathrm{off}}} \right. \nonumber \\
 & \left. \qquad \cdot \left( c_{\mathrm{D,to}}-\frac{\lambda}{\gamma V} l^n_i\right)^{n_{\mathrm{nuc}}+1}\right] \nonumber \\
 &  \cdot \int_{l^n_{i}}^{l_{\mathrm{max}}} \exp \left[-\frac{1}{n_{\mathrm{nuc}}+1}\frac{Vk_{\mathrm{nuc}}}{\lambda k_{\mathrm{off}}} \right. \nonumber \\
 & \qquad \left. \cdot \left( c_{\mathrm{D,to}}-\frac{\lambda}{\gamma V} l^d_i\right)^{n_{\mathrm{nuc}}+1}\right] P_d(l^d_{i}) \dd l^d_i .
\end{align}

In a similar fashion, $P_d(l^d_{i})$ can be expressed as a function of $P_n(l^n_{i-1})$:
\begin{equation}
P_d(l^d_{i}) = \int_{l^n_{\mathrm{min}}}^{l_{\mathrm{max}}} p(l^d_{i}|l^n_{i-1}) P_n(l^n_{i-1}) \, \dd l^n_{i-1} .
\label{eq:distr_rel_2}
\end{equation}
The conditional probability distribution in Eq.~\ref{eq:distr_rel_2} can be obtained from Eq.~\ref{eq:Pcapt_final3} when replacing $T^c$ with the solution of the deterministic relation that connects $l^d_{i}$ and $T^c_{i}$ (Eqs.~\ref{eq:decay} and~\ref{eq:lD_t}):
\begin{align}
l^d_{i} = & -2l^n_{i-1}+2\gamma k_{\mathrm{off}} T^c_{i} + \frac{\gamma V}{\lambda}\left( c_{\mathrm{D,to}}-\frac{k_{\mathrm{off}}}{k_{\mathrm{on}}^{\mathrm{D}}}\right) \nonumber \\
 & - \left[ \frac{\gamma V}{\lambda}\left( c_{\mathrm{D,to}}-\frac{k_{\mathrm{off}}}{k_{\mathrm{on}}^{\mathrm{D}}}\right) -l^n_{i-1}\right] \exp\left(-\frac{\lambda k_{\mathrm{on}}^{\mathrm{D}}}{V}T^c_{i}\right) .
\label{eq:ld1(Tc1)}
\end{align}
Solving this equation for $T^c_{i}$ gives a function $T^c_{i}=g(l^n_{i-1},l^d_{i})$ that involves a LambertW function and can therefore not be expressed in terms of elementary functions. For correct normalisation, the new conditional probability distribution also has to be multiplied by the derivative $\frac{\partial g(l^n_{i-1},l^d_{i})}{\partial l^d_{i}}$. 

The integration limits are the minimal and maximal values of $l^n_{i-1}$, such that the order of events in Fig.~\ref{fig:Ts_and_ls} (i.e. regular oscillations) is still fulfilled for a given $l^d_{i}$. The upper limit turns out to be $l_{\mathrm{max}}$, whereas the lower limit can be found as the solution to the polymer growth equation with $T^c_{i}=T^n_{i-1}$. One finds
\begin{align}
l^n_{\mathrm{min}}= & -\frac{\gamma V}{\lambda k_{\mathrm{on}}^{\mathrm{D}}}\left[ k_{\mathrm{off}} \mathrm{LambertW}\left( -\frac{1}{\gamma V k_{\mathrm{off}}} \right.\right. \nonumber \\
 & \left.\left. \cdot \left( -\gamma V c_{\mathrm{D,to}} k_{\mathrm{on}}^{\mathrm{D}} + \gamma V k_{\mathrm{off}} + \lambda k_{\mathrm{on}}^{\mathrm{D}} l^d_{i}\right) \right.\right. \nonumber \\
& \left.\left. \cdot \exp\left( \frac{c_{\mathrm{D,to}}k_{\mathrm{on}}^{\mathrm{D}}-k_{\mathrm{off}}}{k_{\mathrm{off}}}\right)\right)-c_{\mathrm{D,to}}k_{\mathrm{on}}^{\mathrm{D}}+k_{\mathrm{off}}\right] .
\label{eq:l_n0min}
\end{align}

Putting Eq.~\ref{eq:distr_rel_1} and Eq.~\ref{eq:distr_rel_2} together, one obtains a map from one probability distribution into the next one, half a period later:
\begin{align}
P_n(l^n_{i}) = & \int_{l^n_{i}}^{l_{\mathrm{max}}} \frac{1}{\gamma k_{\mathrm{off}}} P_{\mathrm{nuc}}\left(T^f=\frac{l^d_{i}-l^n_{i}}{\gamma k_{\mathrm{off}}}|l^d_{i}\right) \nonumber \\
& \cdot  \int_{l^n_{\mathrm{min}}}^{l_{\mathrm{max}}} \frac{\partial g(l^n_{i-1},l^d_{i})}{\partial l^d_{i}} P_{\mathrm{cap}}\left(T^c=g(l^n_{i-1},l^d_{i})|l^n_{i-1}\right) \nonumber \\
 & \qquad \cdot  P_n(l^n_{i-1}) \,\dd l^n_{i-1} \,\dd l^d_{i} .
\label{eq:Pln_iteration}
\end{align}

The integrals can be computed numerically\footnote{To numerically compute the integrals, 5000 discrete meshpoints were used over the interval $0\dots L$. After each integration, the distribution is normalised.} and by iterating Eq.~\ref{eq:Pln_iteration}, a steady state probability distribution can be obtained. A typical starting distribution for the iteration would be a delta distribution or a uniform distribution and for typical parameter values, a steady state distribution is reached after 3 to 5 iterations.

This approach is valid, as long as the vast majority of events fall into the scheme depicted in Fig.~\ref{fig:Ts_and_ls}, i.e. as long as $t^d_{i}>t^c_{i+1}$ and $t^d_{i}>t^n_{i+1}$ (regular oscillations -- see Subsec.~\ref{ssec:model_summary}).

A more relevant quantity than $l^n$, the length of a polymer at the time of nucleation on the other side, is the capping length $l^c$, i.e. the amplitude of the oscillation. To obtain the probability distribution for this length, one needs to perform another integration:
\begin{equation}
P_c(l^c) = \int_{l^n_{\mathrm{min}}}^{l^n_{\mathrm{max}}} p(l^c|l^n) P_n(l^n) \, \dd l^n .
\label{eq:P(l_c)}
\end{equation}
The conditional probability distribution can be obtained from Eq.~\ref{eq:Pcapt_final3}, when replacing $T^c$ with the appropriate function of $l^c$ and $l^n$ that can be derived from a similar relation as Eq.~\ref{eq:ld1(Tc1)}. We obtain
\begin{equation}
p(l^c|l^n) = \frac{\partial T^c(l^c,l^n)}{\partial l^c} P_{\mathrm{cap}}(T^c=T^c(l^c,l^n)|l^n) .
\end{equation}
The lower limit of the integration is the same as in the previous integration (Eq.~\ref{eq:l_n0min}). For the upper limit one needs to find $l^n_{i-1}$ such that $l^d_i=0$. This can be found by solving Eq.~\ref{eq:ld1(Tc1)} with $T^c=\frac{1}{\gamma k_{\mathrm{off}}}(l^n-l^c)$ for $l^n$. In case this solution is larger than $l_{\mathrm{max}}$, $l_{\mathrm{max}}$ (Eq.~\ref{eq:lmax}) is the upper limit.

%%%%%%%%%%%%%%%%%%%%%%%%%%%%%%%%%%%%%%%%%%%%%%%%%%%%%%%%%%%%%%%%%%%%%%%%%%%%%

\section{Probability distribution function for the period in the oscillatory state}
\label{app:pdf_T}
\setcounter{equation}{0}

Using the iterative approach described in the previous appendix, the steady state probability distribution function for the lengths of D-polymers at a given time point during regular oscillations can be found. Starting from this distribution, we will compute the probability distribution function for the period of the oscillations in this appendix.

From Fig.~\ref{fig:Ts_and_ls}, half a period $T^h$ (the time between two consecutive nucleations on opposing sides) is found to be $T^h=T^n+T^f$ and therefore $P_h(T^h) = \int p(T^f=T^h-T^n|T^n)p(T^n) \dd T^n$. $T^n$ can be replaced by $T^n=\frac{1}{\gamma k_{\mathrm{off}}}l^n$ and an additional probability distribution $p(l^d|l^n)$ can be introduced in order to obtain an equation involving only distributions that have already been calculated:
\begin{align}
P_h(T^h) = & \int_0^{l^n_{\mathrm{max}}}\int_0^{l^d_{\mathrm{max}}} P_{\mathrm{nuc}}(T^f=T^h-\frac{1}{\gamma k_{\mathrm{off}}}l^n|l^d) \nonumber \\
& \qquad \cdot p(l^d|l^n) \,\dd l^d \, P_n(l^n) \,\dd l^n .
\label{eq:p(T_h)}
\end{align}
$P_{\mathrm{nuc}}(T^f|l^d)$ is given in Eq.~\ref{eq:nuc_distr}, how to obtain $p(l^d|l^n)$ is described in the preceding appendix and a steady state distribution for $P_n(l^n)$ is also derived there (iterating Eq.~\ref{eq:Pln_iteration}). The upper limits of the two integrals are the smaller of $\{ \gamma k_{\mathrm{off}} T^h,l_{\mathrm{max}}\}$ for $l^n_{\mathrm{max}}$ and $l^d_{\mathrm{max}}$ is the equivalent of Eq.~\ref{eq:l_n0min}: $l^d_{\mathrm{max}} = \frac{\gamma V}{\lambda}\left( c_{\mathrm{D,to}}-\frac{k_{\mathrm{off}}}{k_{\mathrm{on}}^{\mathrm{D}}}\right)-\left[\frac{\gamma V}{\lambda}\left( c_{\mathrm{D,to}}-\frac{k_{\mathrm{off}}}{k_{\mathrm{on}}^{\mathrm{D}}}\right)-l^n\right]  \\ \cdot \exp\left(-\frac{\lambda k_{\mathrm{on}}^{\mathrm{D}}}{\gamma k_{\mathrm{off}} V}l^n\right)$.

For the probability distribution function of the period $T=2T^h$, one more integration is needed:
\begin{equation}
P_T(T)=\int_0^T P_h(T^h=T^{h'})P_h(T^h=T-T^{h'}) \,\dd T^{h'} .
\label{eq:p(T)}
\end{equation}

%%%%%%%%%%%%%%%%%%%%%%%%%%%%%%%%%%%%%%%%%%%%%%%%%%%%%%%%%%%%%%%%%%%%%%%%%%%%%

\section{A limiting deterministic map for the case of high cooperativity}
\label{app:underlying_map}
\setcounter{equation}{0}

For the case of deterministic nucleation and stochastic switching, we want to use the deterministic map as shown in Fig.~\ref{fig:map} as a rough guideline for the dynamics of the system. In order to find this underlying map, we need to find a replacement for the parameter $c_{\mathrm{E,th}}$ in the analytical expression of the map (Eqs.~\ref{eq:map_1}--~\ref{eq:map_3}) since the MinE nucleation threshold does not have a meaning if we assume stochastic capping. A natural choice is the E-concentration at the median of the capping cumulative distribution function $F_{\mathrm{cap}}(t)$. In the case of deterministic switching this is a step function in $c_{\mathrm{E}}$ with the step at $c_{\mathrm{E,th}}$. For high $n_{\mathrm{cap}}$, $F_{\mathrm{cap}}(t)$ it is a steep sigmoidal curve. For the vast majority of cases ($k_{\mathrm{cap}}<\frac{\ln(2)}{c_{\mathrm{E},0}^{n_{\mathrm{cap}}} t^{d,\mathrm{E}}}$ and $k_{\mathrm{cap}}>\ln(2)\left( c_{\mathrm{E},0}^{n_{\mathrm{cap}}} t^{d,\mathrm{E}}+\frac{c_{\mathrm{E,to}}^{n_{\mathrm{cap}}+1}-c_{\mathrm{E},0}^{n_{\mathrm{cap}}+1}}{A(n_{\mathrm{cap}}+1)}\right)^{-1}$), the median capping time $t_M$ lies within $t^{d,\mathrm{E}}\le t_M \le t^d$. With the integration of Eq.~\ref{eq:Pcapt_final3} (and $l^n=\frac{\gamma V}{\lambda}(c_{\mathrm{D,to}}-c_{\mathrm{D,th}})$) one finds the median time to be  $t_M=t^{d,\mathrm{E}}+\frac{1}{A}  \Bigl( -c_{\mathrm{E},0} + \Bigl[ c_{\mathrm{E},0}^{n_{\mathrm{cap}}+1}  -\frac{A(n_{\mathrm{cap}}+1)}{k_{\mathrm{cap}}}\Bigl( k_{\mathrm{cap}}c_{\mathrm{E},0}^{n_{\mathrm{cap}}} t^{d,\mathrm{E}} - \ln(2)\Bigr) \Bigr]^{\frac{1}{n_{\mathrm{cap}}+1}}\Bigr)$, which then leads to
\begin{align}
c_{\mathrm{E}}(t_M) = & \Bigl[ c_{\mathrm{E},0}^{n_{\mathrm{cap}}+1}-\frac{A(n_{\mathrm{cap}}+1)}{k_{\mathrm{cap}}}  \nonumber \\
 &  \cdot \left( k_{\mathrm{cap}}c_{\mathrm{E},0}^{n_{\mathrm{cap}}}t^{d,\mathrm{E}} - \ln(2)\right)\Bigr]^{\frac{1}{n_{\mathrm{cap}}+1}} .
\end{align}

We use this E-concentration as a replacement of $c_{\mathrm{E,th}}$ in Eqs.~\ref{eq:map_1}--\ref{eq:map_3} and plot the resulting maps in Fig.~\ref{fig:episodes1:map}.

%%%%%%%%%%%%%%%%%%%%%%%%%%%%%%%%%%%%%%%%%%%%%%%%%%%%%%%%%%%%%%%%%%%%%%%%%%%%%

\bibliographystyle{apsrev}
\bibliography{../../bibliography_Min}

\begin{thebibliography}{33}
\expandafter\ifx\csname natexlab\endcsname\relax\def\natexlab#1{#1}\fi
\expandafter\ifx\csname bibnamefont\endcsname\relax
  \def\bibnamefont#1{#1}\fi
\expandafter\ifx\csname bibfnamefont\endcsname\relax
  \def\bibfnamefont#1{#1}\fi
\expandafter\ifx\csname citenamefont\endcsname\relax
  \def\citenamefont#1{#1}\fi
\expandafter\ifx\csname url\endcsname\relax
  \def\url#1{\texttt{#1}}\fi
\expandafter\ifx\csname urlprefix\endcsname\relax\def\urlprefix{URL }\fi
\providecommand{\bibinfo}[2]{#2}
\providecommand{\eprint}[2][]{\url{#2}}

\bibitem[{\citenamefont{Lutkenhaus}(2008)}]{Lutkenhaus2009}
\bibinfo{author}{\bibfnamefont{J.}~\bibnamefont{Lutkenhaus}},
  \bibinfo{journal}{Adv. Exp. Med. Biol.} \textbf{\bibinfo{volume}{641}},
  \bibinfo{pages}{49} (\bibinfo{year}{2008}).

\bibitem[{\citenamefont{Meinhardt and de~Boer}(2001)}]{Meinhardt2001}
\bibinfo{author}{\bibfnamefont{H.}~\bibnamefont{Meinhardt}} \bibnamefont{and}
  \bibinfo{author}{\bibfnamefont{P.~A.} \bibnamefont{de~Boer}},
  \bibinfo{journal}{Proc Natl Acad Sci U S A} \textbf{\bibinfo{volume}{98}},
  \bibinfo{pages}{14202} (\bibinfo{year}{2001}).

\bibitem[{\citenamefont{Meacci and Kruse}(2005)}]{Meacci2005}
\bibinfo{author}{\bibfnamefont{G.}~\bibnamefont{Meacci}} \bibnamefont{and}
  \bibinfo{author}{\bibfnamefont{K.}~\bibnamefont{Kruse}},
  \bibinfo{journal}{Phys Biol} \textbf{\bibinfo{volume}{2}},
  \bibinfo{pages}{89} (\bibinfo{year}{2005}).

\bibitem[{\citenamefont{Huang et~al.}(2003)\citenamefont{Huang, Meir, and
  Wingreen}}]{Huang2003}
\bibinfo{author}{\bibfnamefont{K.~C.} \bibnamefont{Huang}},
  \bibinfo{author}{\bibfnamefont{Y.}~\bibnamefont{Meir}}, \bibnamefont{and}
  \bibinfo{author}{\bibfnamefont{N.~S.} \bibnamefont{Wingreen}},
  \bibinfo{journal}{Proc Natl Acad Sci U S A} \textbf{\bibinfo{volume}{100}},
  \bibinfo{pages}{12724} (\bibinfo{year}{2003}).

\bibitem[{\citenamefont{Howard et~al.}(2001)\citenamefont{Howard, Rutenberg,
  and de~Vet}}]{Howard2001}
\bibinfo{author}{\bibfnamefont{M.}~\bibnamefont{Howard}},
  \bibinfo{author}{\bibfnamefont{A.~D.} \bibnamefont{Rutenberg}},
  \bibnamefont{and} \bibinfo{author}{\bibfnamefont{S.}~\bibnamefont{de~Vet}},
  \bibinfo{journal}{Phys Rev Lett} \textbf{\bibinfo{volume}{87}},
  \bibinfo{pages}{278102} (\bibinfo{year}{2001}).

\bibitem[{\citenamefont{Kerr et~al.}(2006)\citenamefont{Kerr, Levine,
  Sejnowski, and Rappel}}]{Kerr2006}
\bibinfo{author}{\bibfnamefont{R.}~\bibnamefont{Kerr}},
  \bibinfo{author}{\bibfnamefont{H.}~\bibnamefont{Levine}},
  \bibinfo{author}{\bibfnamefont{T.}~\bibnamefont{Sejnowski}},
  \bibnamefont{and} \bibinfo{author}{\bibfnamefont{W.}~\bibnamefont{Rappel}},
  \bibinfo{journal}{Proc. Natl. Acad. Sci. U.S.A.}
  \textbf{\bibinfo{volume}{103}}, \bibinfo{pages}{347} (\bibinfo{year}{2006}).

\bibitem[{\citenamefont{Fange and Elf}(2006)}]{Fange2006}
\bibinfo{author}{\bibfnamefont{D.}~\bibnamefont{Fange}} \bibnamefont{and}
  \bibinfo{author}{\bibfnamefont{J.}~\bibnamefont{Elf}}, \bibinfo{journal}{PLoS
  Comput. Biol.} \textbf{\bibinfo{volume}{2}}, \bibinfo{pages}{e80}
  (\bibinfo{year}{2006}).

\bibitem[{\citenamefont{Kruse}(2002)}]{Kruse2002}
\bibinfo{author}{\bibfnamefont{K.}~\bibnamefont{Kruse}},
  \bibinfo{journal}{Biophys. J.} \textbf{\bibinfo{volume}{82}},
  \bibinfo{pages}{618} (\bibinfo{year}{2002}).

\bibitem[{\citenamefont{Drew et~al.}(2005)\citenamefont{Drew, Osborn, and
  Rothfield}}]{Drew2005}
\bibinfo{author}{\bibfnamefont{D.~A.} \bibnamefont{Drew}},
  \bibinfo{author}{\bibfnamefont{M.~J.} \bibnamefont{Osborn}},
  \bibnamefont{and} \bibinfo{author}{\bibfnamefont{L.~I.}
  \bibnamefont{Rothfield}}, \bibinfo{journal}{Proc Natl Acad Sci U S A}
  \textbf{\bibinfo{volume}{102}}, \bibinfo{pages}{6114} (\bibinfo{year}{2005}).

\bibitem[{\citenamefont{Pavin et~al.}(2006)\citenamefont{Pavin, Paljetak, and
  Krstic}}]{Pavin2006}
\bibinfo{author}{\bibfnamefont{N.}~\bibnamefont{Pavin}},
  \bibinfo{author}{\bibfnamefont{H.~C.} \bibnamefont{Paljetak}},
  \bibnamefont{and} \bibinfo{author}{\bibfnamefont{V.}~\bibnamefont{Krstic}},
  \bibinfo{journal}{Phys Rev E Stat Nonlin Soft Matter Phys}
  \textbf{\bibinfo{volume}{73}}, \bibinfo{pages}{021904}
  (\bibinfo{year}{2006}).

\bibitem[{\citenamefont{Cytrynbaum and Marshall}(2007)}]{Cytrynbaum2007}
\bibinfo{author}{\bibfnamefont{E.~N.} \bibnamefont{Cytrynbaum}}
  \bibnamefont{and} \bibinfo{author}{\bibfnamefont{B.~D.~L.}
  \bibnamefont{Marshall}}, \bibinfo{journal}{Biophys J}
  \textbf{\bibinfo{volume}{93}}, \bibinfo{pages}{1134} (\bibinfo{year}{2007}).

\bibitem[{\citenamefont{Tostevin and Howard}(2006)}]{Tostevin2006}
\bibinfo{author}{\bibfnamefont{F.}~\bibnamefont{Tostevin}} \bibnamefont{and}
  \bibinfo{author}{\bibfnamefont{M.}~\bibnamefont{Howard}},
  \bibinfo{journal}{Phys Biol} \textbf{\bibinfo{volume}{3}}, \bibinfo{pages}{1}
  (\bibinfo{year}{2006}).

\bibitem[{\citenamefont{Hu et~al.}(2002)\citenamefont{Hu, Gogol, and
  Lutkenhaus}}]{Hu2002}
\bibinfo{author}{\bibfnamefont{Z.}~\bibnamefont{Hu}},
  \bibinfo{author}{\bibfnamefont{E.~P.} \bibnamefont{Gogol}}, \bibnamefont{and}
  \bibinfo{author}{\bibfnamefont{J.}~\bibnamefont{Lutkenhaus}},
  \bibinfo{journal}{Proc Natl Acad Sci U S A} \textbf{\bibinfo{volume}{99}},
  \bibinfo{pages}{6761} (\bibinfo{year}{2002}).

\bibitem[{\citenamefont{Suefuji et~al.}(2002)\citenamefont{Suefuji, Valluzzi,
  and RayChaudhuri}}]{Suefuji2002}
\bibinfo{author}{\bibfnamefont{K.}~\bibnamefont{Suefuji}},
  \bibinfo{author}{\bibfnamefont{R.}~\bibnamefont{Valluzzi}}, \bibnamefont{and}
  \bibinfo{author}{\bibfnamefont{D.}~\bibnamefont{RayChaudhuri}},
  \bibinfo{journal}{Proc Natl Acad Sci U S A} \textbf{\bibinfo{volume}{99}},
  \bibinfo{pages}{16776} (\bibinfo{year}{2002}).

\bibitem[{\citenamefont{Shih et~al.}(2003)\citenamefont{Shih, Le, and
  Rothfield}}]{Shih2003}
\bibinfo{author}{\bibfnamefont{Y.-L.} \bibnamefont{Shih}},
  \bibinfo{author}{\bibfnamefont{T.}~\bibnamefont{Le}}, \bibnamefont{and}
  \bibinfo{author}{\bibfnamefont{L.}~\bibnamefont{Rothfield}},
  \bibinfo{journal}{Proc Natl Acad Sci U S A} \textbf{\bibinfo{volume}{100}},
  \bibinfo{pages}{7865} (\bibinfo{year}{2003}).

\bibitem[{\citenamefont{Szeto et~al.}(2005)\citenamefont{Szeto, Eng, Acharya,
  Rigden, and Dillon}}]{Szeto2005}
\bibinfo{author}{\bibfnamefont{J.}~\bibnamefont{Szeto}},
  \bibinfo{author}{\bibfnamefont{N.~F.} \bibnamefont{Eng}},
  \bibinfo{author}{\bibfnamefont{S.}~\bibnamefont{Acharya}},
  \bibinfo{author}{\bibfnamefont{M.~D.} \bibnamefont{Rigden}},
  \bibnamefont{and} \bibinfo{author}{\bibfnamefont{J.-A.~R.}
  \bibnamefont{Dillon}}, \bibinfo{journal}{Res Microbiol}
  \textbf{\bibinfo{volume}{156}}, \bibinfo{pages}{17} (\bibinfo{year}{2005}).

\bibitem[{\citenamefont{Downing}()}]{Benjamin}
\bibinfo{author}{\bibfnamefont{B.}~\bibnamefont{Downing}},
  \bibinfo{howpublished}{personal communication}.

\bibitem[{\citenamefont{Yu and Margolin}(1999)}]{Yu1999}
\bibinfo{author}{\bibfnamefont{X.~C.} \bibnamefont{Yu}} \bibnamefont{and}
  \bibinfo{author}{\bibfnamefont{W.}~\bibnamefont{Margolin}},
  \bibinfo{journal}{Mol Microbiol} \textbf{\bibinfo{volume}{32}},
  \bibinfo{pages}{315} (\bibinfo{year}{1999}).

\bibitem[{\citenamefont{de~Boer et~al.}(1991)\citenamefont{de~Boer, Crossley,
  Hand, and Rothfield}}]{deBoer1991}
\bibinfo{author}{\bibfnamefont{P.~A.} \bibnamefont{de~Boer}},
  \bibinfo{author}{\bibfnamefont{R.~E.} \bibnamefont{Crossley}},
  \bibinfo{author}{\bibfnamefont{A.~R.} \bibnamefont{Hand}}, \bibnamefont{and}
  \bibinfo{author}{\bibfnamefont{L.~I.} \bibnamefont{Rothfield}},
  \bibinfo{journal}{EMBO J.} \textbf{\bibinfo{volume}{10}},
  \bibinfo{pages}{4371} (\bibinfo{year}{1991}).

\bibitem[{\citenamefont{Hu and Lutkenhaus}(2001)}]{Hu2001}
\bibinfo{author}{\bibfnamefont{Z.}~\bibnamefont{Hu}} \bibnamefont{and}
  \bibinfo{author}{\bibfnamefont{J.}~\bibnamefont{Lutkenhaus}},
  \bibinfo{journal}{Mol. Cell} \textbf{\bibinfo{volume}{7}},
  \bibinfo{pages}{1337} (\bibinfo{year}{2001}).

\bibitem[{\citenamefont{Raskin and de~Boer}(1999{\natexlab{a}})}]{Raskin1999b}
\bibinfo{author}{\bibfnamefont{D.}~\bibnamefont{Raskin}} \bibnamefont{and}
  \bibinfo{author}{\bibfnamefont{P.}~\bibnamefont{de~Boer}},
  \bibinfo{journal}{J. Bacteriol.} \textbf{\bibinfo{volume}{181}},
  \bibinfo{pages}{6419} (\bibinfo{year}{1999}{\natexlab{a}}).

\bibitem[{\citenamefont{Hu et~al.}(1999)\citenamefont{Hu, Mukherjee, Pichoff,
  and Lutkenhaus}}]{Hu1999}
\bibinfo{author}{\bibfnamefont{Z.}~\bibnamefont{Hu}},
  \bibinfo{author}{\bibfnamefont{A.}~\bibnamefont{Mukherjee}},
  \bibinfo{author}{\bibfnamefont{S.}~\bibnamefont{Pichoff}}, \bibnamefont{and}
  \bibinfo{author}{\bibfnamefont{J.}~\bibnamefont{Lutkenhaus}},
  \bibinfo{journal}{Proc. Natl. Acad. Sci. U.S.A.}
  \textbf{\bibinfo{volume}{96}}, \bibinfo{pages}{14819} (\bibinfo{year}{1999}).

\bibitem[{\citenamefont{Raskin and de~Boer}(1999{\natexlab{b}})}]{Raskin1999}
\bibinfo{author}{\bibfnamefont{D.~M.} \bibnamefont{Raskin}} \bibnamefont{and}
  \bibinfo{author}{\bibfnamefont{P.~A.} \bibnamefont{de~Boer}},
  \bibinfo{journal}{Proc Natl Acad Sci U S A} \textbf{\bibinfo{volume}{96}},
  \bibinfo{pages}{4971} (\bibinfo{year}{1999}{\natexlab{b}}).

\bibitem[{\citenamefont{Mileykovskaya et~al.}(2009)\citenamefont{Mileykovskaya,
  Ryan, Mo, Lin, Khalaf, Dowhan, and Garrett}}]{Mileykovskaya2009}
\bibinfo{author}{\bibfnamefont{E.}~\bibnamefont{Mileykovskaya}},
  \bibinfo{author}{\bibfnamefont{A.~C.} \bibnamefont{Ryan}},
  \bibinfo{author}{\bibfnamefont{X.}~\bibnamefont{Mo}},
  \bibinfo{author}{\bibfnamefont{C.~C.} \bibnamefont{Lin}},
  \bibinfo{author}{\bibfnamefont{K.~I.} \bibnamefont{Khalaf}},
  \bibinfo{author}{\bibfnamefont{W.}~\bibnamefont{Dowhan}}, \bibnamefont{and}
  \bibinfo{author}{\bibfnamefont{T.~A.} \bibnamefont{Garrett}},
  \bibinfo{journal}{J. Biol. Chem.} \textbf{\bibinfo{volume}{284}},
  \bibinfo{pages}{2990} (\bibinfo{year}{2009}).

\bibitem[{\citenamefont{Mazor et~al.}(2008)\citenamefont{Mazor, Regev,
  Mileykovskaya, Margolin, Dowhan, and Fishov}}]{Mazor2008}
\bibinfo{author}{\bibfnamefont{S.}~\bibnamefont{Mazor}},
  \bibinfo{author}{\bibfnamefont{T.}~\bibnamefont{Regev}},
  \bibinfo{author}{\bibfnamefont{E.}~\bibnamefont{Mileykovskaya}},
  \bibinfo{author}{\bibfnamefont{W.}~\bibnamefont{Margolin}},
  \bibinfo{author}{\bibfnamefont{W.}~\bibnamefont{Dowhan}}, \bibnamefont{and}
  \bibinfo{author}{\bibfnamefont{I.}~\bibnamefont{Fishov}},
  \bibinfo{journal}{Biochim Biophys Acta} \textbf{\bibinfo{volume}{1778}},
  \bibinfo{pages}{2505} (\bibinfo{year}{2008}).

\bibitem[{\citenamefont{Touhami et~al.}(2006)\citenamefont{Touhami, Jericho,
  and Rutenberg}}]{Touhami2006}
\bibinfo{author}{\bibfnamefont{A.}~\bibnamefont{Touhami}},
  \bibinfo{author}{\bibfnamefont{M.}~\bibnamefont{Jericho}}, \bibnamefont{and}
  \bibinfo{author}{\bibfnamefont{A.~D.} \bibnamefont{Rutenberg}},
  \bibinfo{journal}{J Bacteriol} \textbf{\bibinfo{volume}{188}},
  \bibinfo{pages}{7661} (\bibinfo{year}{2006}).

\bibitem[{\citenamefont{Meacci et~al.}(2006)\citenamefont{Meacci, Ries,
  Fischer-Friedrich, Kahya, Schwille, and Kruse}}]{Meacci2006}
\bibinfo{author}{\bibfnamefont{G.}~\bibnamefont{Meacci}},
  \bibinfo{author}{\bibfnamefont{J.}~\bibnamefont{Ries}},
  \bibinfo{author}{\bibfnamefont{E.}~\bibnamefont{Fischer-Friedrich}},
  \bibinfo{author}{\bibfnamefont{N.}~\bibnamefont{Kahya}},
  \bibinfo{author}{\bibfnamefont{P.}~\bibnamefont{Schwille}}, \bibnamefont{and}
  \bibinfo{author}{\bibfnamefont{K.}~\bibnamefont{Kruse}},
  \bibinfo{journal}{Phys Biol} \textbf{\bibinfo{volume}{3}},
  \bibinfo{pages}{255} (\bibinfo{year}{2006}).

\bibitem[{\citenamefont{Champneys and di~Bernardo}(2008)}]{Champneys:2008}
\bibinfo{author}{\bibfnamefont{A.~R.} \bibnamefont{Champneys}}
  \bibnamefont{and}
  \bibinfo{author}{\bibfnamefont{M.}~\bibnamefont{di~Bernardo}},
  \bibinfo{journal}{Scholarpedia} \textbf{\bibinfo{volume}{3}},
  \bibinfo{pages}{4041} (\bibinfo{year}{2008}).

\bibitem[{\citenamefont{Mileykovskaya et~al.}(2003)\citenamefont{Mileykovskaya,
  Fishov, Fu, Corbin, Margolin, and Dowhan}}]{Mileykovskaya2003}
\bibinfo{author}{\bibfnamefont{E.}~\bibnamefont{Mileykovskaya}},
  \bibinfo{author}{\bibfnamefont{I.}~\bibnamefont{Fishov}},
  \bibinfo{author}{\bibfnamefont{X.}~\bibnamefont{Fu}},
  \bibinfo{author}{\bibfnamefont{B.}~\bibnamefont{Corbin}},
  \bibinfo{author}{\bibfnamefont{W.}~\bibnamefont{Margolin}}, \bibnamefont{and}
  \bibinfo{author}{\bibfnamefont{W.}~\bibnamefont{Dowhan}},
  \bibinfo{journal}{J. Biol. Chem.} \textbf{\bibinfo{volume}{278}},
  \bibinfo{pages}{22193} (\bibinfo{year}{2003}).

\bibitem[{\citenamefont{Oosawa and Asakura}(1975)}]{Oosawa1975}
\bibinfo{author}{\bibfnamefont{F.}~\bibnamefont{Oosawa}} \bibnamefont{and}
  \bibinfo{author}{\bibfnamefont{S.}~\bibnamefont{Asakura}},
  \emph{\bibinfo{title}{Thermodynamics of the Polymerization of Protein
  (Molecular Biology)}} (\bibinfo{publisher}{Academic Press Inc},
  \bibinfo{year}{1975}).

\bibitem[{\citenamefont{Shih et~al.}(2002)\citenamefont{Shih, Fu, King, Le, and
  Rothfield}}]{Shih2002}
\bibinfo{author}{\bibfnamefont{Y.-L.} \bibnamefont{Shih}},
  \bibinfo{author}{\bibfnamefont{X.}~\bibnamefont{Fu}},
  \bibinfo{author}{\bibfnamefont{G.~F.} \bibnamefont{King}},
  \bibinfo{author}{\bibfnamefont{T.}~\bibnamefont{Le}}, \bibnamefont{and}
  \bibinfo{author}{\bibfnamefont{L.}~\bibnamefont{Rothfield}},
  \bibinfo{journal}{EMBO J} \textbf{\bibinfo{volume}{21}},
  \bibinfo{pages}{3347} (\bibinfo{year}{2002}).

\bibitem[{\citenamefont{Hale et~al.}(2001)\citenamefont{Hale, Meinhardt, and
  de~Boer}}]{Hale2001}
\bibinfo{author}{\bibfnamefont{C.}~\bibnamefont{Hale}},
  \bibinfo{author}{\bibfnamefont{H.}~\bibnamefont{Meinhardt}},
  \bibnamefont{and} \bibinfo{author}{\bibfnamefont{P.}~\bibnamefont{de~Boer}},
  \bibinfo{journal}{EMBO J.} \textbf{\bibinfo{volume}{20}},
  \bibinfo{pages}{1563} (\bibinfo{year}{2001}).

\bibitem[{\citenamefont{de~Boer et~al.}(1989)\citenamefont{de~Boer, Crossley,
  and Rothfield}}]{deBoer1989}
\bibinfo{author}{\bibfnamefont{P.}~\bibnamefont{de~Boer}},
  \bibinfo{author}{\bibfnamefont{R.}~\bibnamefont{Crossley}}, \bibnamefont{and}
  \bibinfo{author}{\bibfnamefont{L.}~\bibnamefont{Rothfield}},
  \bibinfo{journal}{Cell} \textbf{\bibinfo{volume}{56}}, \bibinfo{pages}{641}
  (\bibinfo{year}{1989}).

\end{thebibliography}

\end{document}